\documentclass[]{achemso}

\usepackage{amsmath}        
\usepackage{commath}        
\usepackage{graphicx}       
\graphicspath{{figures/}}   
\usepackage{subcaption}     
\usepackage{booktabs}       
\usepackage[table]{xcolor}  
\usepackage{makecell}       
\usepackage{etoolbox}       
\apptocmd{\thebibliography}{\raggedright}{}{}
\usepackage{glossaries}     

\newacronym{DFT}{DFT}{Density Functional Theory}
\newacronym{VASP}{VASP}{the Vienna Ab-initio Simulation Package}
\newacronym{GASpy}{GASpy}{the Generalized Adsorption Simulator for Python}
\newacronym{rPBE}{rPBE}{revised Perdew-Burke-Ernzerhof}
\newacronym{ASE}{ASE}{the Atomic Simulation Environment}
\newacronym{NEB}{NEB}{nudged elastic band}
\newacronym{GP}{GP}{Gaussian Process}
\newacronym{UQ}{UQ}{uncertainty quantification}
\newacronym{MAE}{MAE}{Mean Absolute Error}
\newacronym{RMSE}{RMSE}{Root Mean Squared Error}
\newacronym{MDAE}{MDAE}{Median Absolute Error}
\newacronym{MARPD}{MARPD}{Mean Absolute Relative Percent Difference}
\newacronym{R2}{R\textsuperscript{2}}{R\textsuperscript{2} correlation coefficient}
\newacronym{Cv}{C\textsubscript{v}}{coefficient of variation}
\newacronym{NN}{NN}{crystal graph convolutional Neural Network}
\newacronym{dropout}{Dropout NN}{Dropout Neural Networks}
\newacronym{BNN}{BNN}{Bayesian Neural Network}
\newacronym{MLE}{MLE}{Maximum Likelihood Estimation}
\newacronym{CFGP}{CFGP}{Convolution-Fed Gaussian Process}
\newacronym{dNN}{NN$\Delta$NN}{in-series NNs}
\newacronym{dGP}{GP$_{NN-\mu}$}{GP with NN mean}
\newacronym{ML}{ML}{Machine Learning}
\newacronym{eV}{eV}{electron volts}
\newacronym{NLL}{NLL}{negative log-likelihood}
\newacronym{BEEF}{BEEF-vdW}{Bayesian Error Estimation Functional with van der Waals correlation}

\title{Methods for comparing uncertainty quantifications for material property predictions}
\author{Kevin Tran}
\affiliation{Chemical Engineering Department, Carnegie Mellon University, Pittsburgh, PA 15217}
\altaffiliation{These authors contributed equally to this work}
\author{Willie Neiswanger}
\affiliation{Machine Learning Department, Carnegie Mellon University, Pittsburgh, PA 15217}
\altaffiliation{These authors contributed equally to this work}
\author{Junwoong Yoon}
\affiliation{Chemical Engineering Department, Carnegie Mellon University, Pittsburgh, PA 15217}
\author{Qingyang Zhang}
\affiliation{Chemical Engineering Department, Carnegie Mellon University, Pittsburgh, PA 15217}
\author{Eric Xing}
\affiliation{Machine Learning Department, Carnegie Mellon University, Pittsburgh, PA 15217}
\author{Zachary W. Ulissi}
\affiliation{Chemical Engineering Department, Carnegie Mellon University, Pittsburgh, PA 15217}
\email{zulissi@andrew.cmu.edu}

\begin{document}

\begin{abstract}
    Data science and informatics tools have been proliferating recently within the computational materials science and catalysis fields.
    This proliferation has spurned the creation of various frameworks for automated materials screening, discovery, and design.
    Underpinning these frameworks are surrogate models with uncertainty estimates on their predictions.
    These uncertainty estimates are instrumental for determining which materials to screen next, but the computational catalysis field does not yet have a standard procedure for judging the quality of such uncertainty estimates.
    Here we present a suite of figures and performance metrics derived from the machine learning community that can be used to judge the quality of such uncertainty estimates.
    This suite probes the accuracy, calibration, and sharpness of a model quantitatively.
    We then show a case study where we judge various methods for predicting density-functional-theory-calculated adsorption energies.
    Of the methods studied here, we find that the best performer is a model where a convolutional neural network is used to supply features to a Gaussian process regressor, which then makes predictions of adsorption energies along with corresponding uncertainty estimates.
\end{abstract}


\section{Introduction}

The fields of catalysis and materials science are burgeoning with methods to screen, design, and understand materials.\cite{Medford2018, Gu2019, Schleder2019, Alberi2019}
This research has spurned the creation of \gls{ML} models to predict various material properties.
Unfortunately, the design spaces for these models are sometimes too large and intractable to sample completely.
These under-sampling issues can limit the training data and therefore the predictive power of the models.
It would be helpful to have an \gls{UQ} for a model so that we know when to trust the predictions and when not to.
More specifically:  \gls{UQ} would enable various online, active frameworks for materials discovery and design (e.g., active learning,\cite{Settles2012} online active learning,\cite{Chu2011} Bayesian optimization,\cite{Frazier2018} active search,\cite{Garnett2012} or goal oriented design of experiments\cite{Kandasamy}).

Such active frameworks have already been used successfully in the field of catalysis and materials informatics.
For example: \citet{Peterson2016} has used a neural network to perform online active learning of \gls{NEB} calculations, reducing the number of force calls by an order of magnitude.
\citet{Torres2018} have also used online active learning to accelerate \gls{NEB} calculations, but they used a \gls{GP} model instead of a neural network.
\citet{Jinnouchi2019} have used online active learning to accelerate molecular dynamics simulations.
These methods are all underpinned by models with \gls{UQ}, which have garnered increasing attention.\cite{Peterson2017, Musil2019}

The goal of \gls{UQ} is to quantify accurately the likelihood of outcomes associated with a predicted quantity.
For example, given an input for which we wish to make a prediction, a predictive \gls{UQ} method might return a confidence interval that aims to capture the true outcome a specified percentage of the time or might return a probability distribution over possible outcomes.
Performance metrics for predictive \gls{UQ} methods aim to assess how well a given quantification of the probabilities of potential true outcomes adheres to a set of observations of these outcomes.
Some of the performance metrics for predictive \gls{UQ} are agnostic to prediction performance---they provide an assessment of the uncertainty independent of the predictive accuracy (i.e.\ a method can predict badly, but could still accurately quantify its own uncertainty).

\begin{figure}
    \centering
    \includegraphics[width=0.75\textwidth]{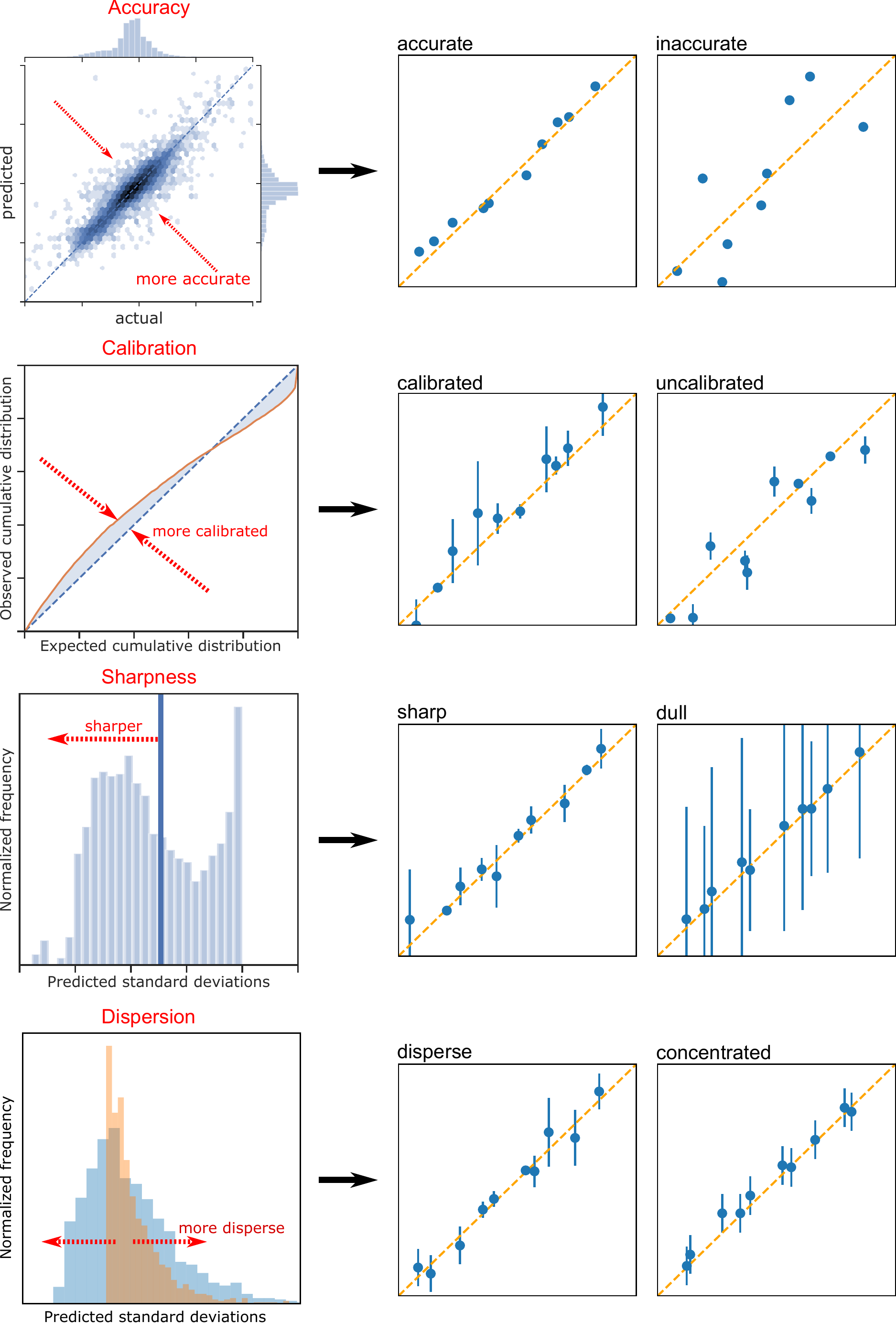}
    \caption{Overview of proposed procedure for judging the quality of models with uncertainty estimates.
    First and foremost, the models should be accurate.
    Second, the models should be ``calibrated'', which means that their uncertainty estimates should be comparable with their residuals.
    Third, the models should be ``sharp'', which means that their uncertainty estimates should be low.
    Lastly, the models should be ``disperse'', which means that the distribution of the uncertainty estimates should be wide.
    This study demonstrates how to visualize and quantify these characteristics so that different methods of \gls{UQ} can be compared objectively.}\label{fig:overview}
\end{figure}

We have seen few\cite{Janet2019, Scalia2019} comparisons of different methods for \gls{UQ} within the field of catalysis and materials informatics.
Here we examine a protocol\cite{Kuleshov2018, Levi2020} for comparing the performance of different modeling and \gls{UQ} methods (Figure~\ref{fig:overview}).
We then illustrate the protocol on a case study where we compare various models' abilities to predict \gls{DFT} calculated adsorption energies.
We also offer anecdotal insights from our case study.
We acknowledge that such insights may not be transferable to other applications, but we find value in sharing them so that others can build their own intuition.

\section{Methods}

\subsection{Dataset information}

All regressions in this paper were performed using a dataset of 47,279 \gls{DFT} calculated adsorption energies created with \gls{GASpy}\cite{Tran2018, Tran2018a}.
Within this dataset, there were 52 different elements within the 1,952 bulk structures used as bases for the adsorption surfaces.
The 61 bulk structures that contained one element encompassed 5,844 of the adsorption calculations; the 1,057 bulk structures that contained two elements encompassed 31,651 of the calculations; the 774 bulk structures that contained three elements encompassed 9,139 of the calculations; and the 60 bulk structures that contained four or five elements encompassed 645 of the calculations.
The dataset also comprised 9,102 symmetrically distinct surfaces and 29,843 distinct coordination environments (as defined by the surface and the adsorbate neighbors).
Lastly, the dataset comprised 21,269 H adsorption energies; 18,437 CO adsorption energies; 3,464 OH adsorption energies; 2,515 O adsorption energies; and 1,594 N adsorption energies.

\gls{GASpy} performed all \gls{DFT} calculations using \gls{VASP}\cite{Kresse1993, Kresse1994, Kresse1996, Kresse1996a} version 5.4 implemented in \gls{ASE}\cite{HjorthLarsen2017}.
The \gls{rPBE} functionals\cite{Hammer1999} were used along with \gls{VASP}'s pseudopotentials, and no spin magnetism or dispersion corrections were used.
Bulk relaxations were performed with a $10\times10\times10$ k-point grid and a 500 \gls{eV} cutoff, and only isotropic relaxation were allowed during this bulk relaxation.
Slab relaxations were performed with k-point grids of $4\times4\times1$ and a 350 \gls{eV} cutoff.
Slabs were replicated in the X/Y directions so that each cell was at least 4.5 \AA{} wide, which reduces adsorbate self-interaction.
Slabs were also replicated in the Z direction until they were at least 7 \AA{} thick, and at least 20 \AA{} of vacuum was included in between slabs.
The bottom layers of each slab were fixed and defined as those atoms more than 3 \AA{} from the top of the surface in the scaled Z direction.

To split the data into train/validate/test sets, we enumerated all adsorption energies on monometallic slabs and added them to the training set manually. 
We did this because some of the regression methods in this paper use a featurization that contains our monometallic adsorption energy data\cite{Tran2018}, and so having the monometallic adsorption energies pre-allocated in the training set prevented any information leakage between the training set and validation/test sets.
After this allocation, we performed a 64/14/20 train/validate/test split that was stratified\cite{Thompson2012} by adsorbate.
We then used the validation set's results to tune various hyperparameters manually.
After tuning, we calculated the test set results and present them in this paper exclusively.
Note that the test results were obtained using models that were trained only using the training set, not the validation set.
This is acceptable because we only seek to compare methods here, not to optimize them.

Note that random splits such as this may yield overly optimistic model results.
If a model created with the training set is meant to make extrapolative predictions in feature domains outside of the training set, then it may be appropriate to use a train/validate/test split using k-means clustering\cite{Meredig2018} rather than random splitting.
If the model is meant to be used in an online and iterative fashion, then it may be appropriate to use a time-series split\cite{Hyndman2014}.
If the model is meant to be used to interpolate within a given feature space, then the basic random split may be appropriate.
We chose to use a basic random split in this work to simplify the results for illustrative purposes.
Future work for different applications should use splitting methods that align with the intended use of the models to be generated.

\subsection{Regression methods}

\begin{figure}
    \centering
    \includegraphics[width=0.8\textwidth]{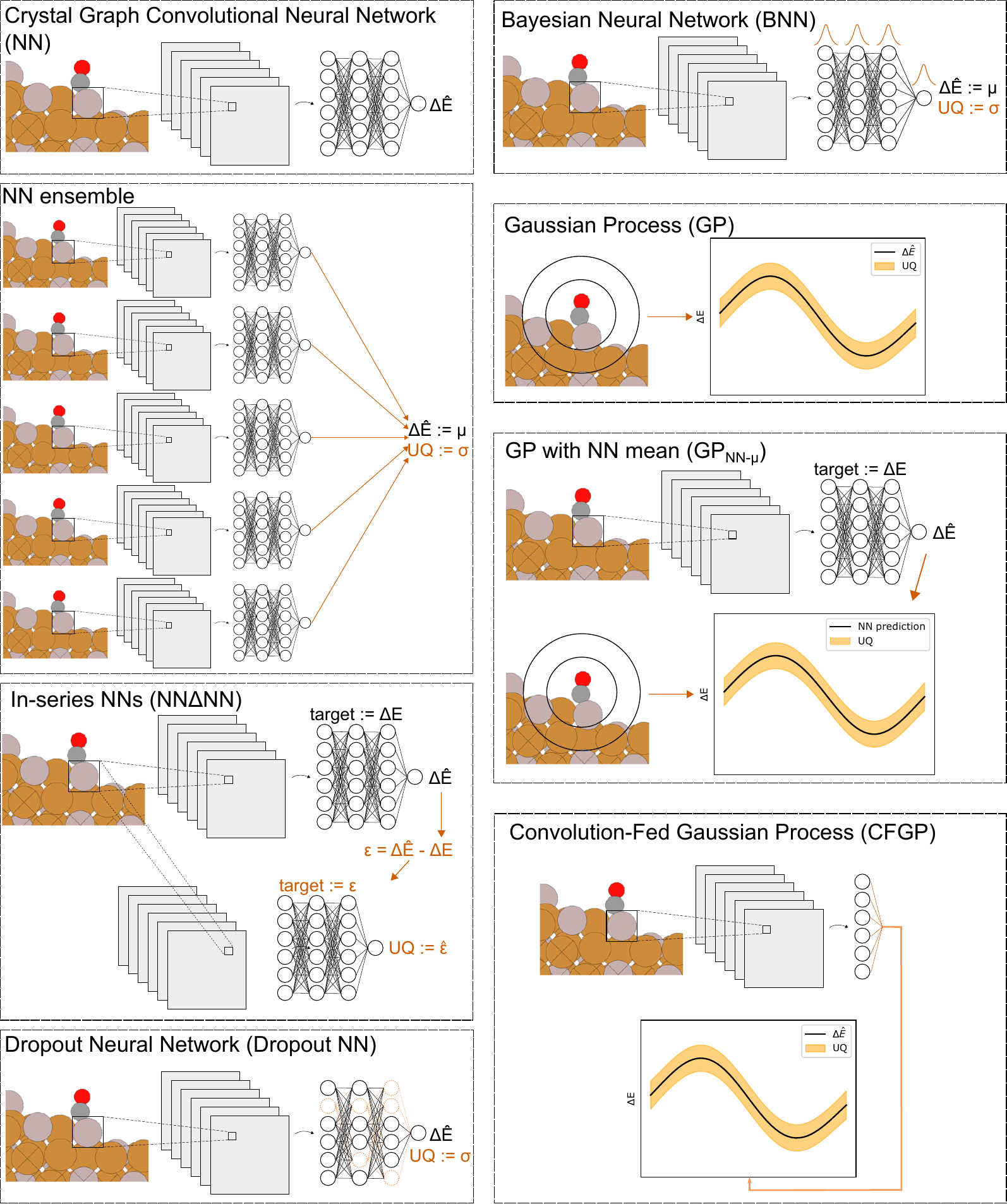}
    \caption{Overview of the various \gls{UQ} methods we investigated in this study.
    $\Delta E$ represents \gls{DFT}-calculated adsorption energies;
    $\Delta \hat{E}$ represents \gls{ML}-predicted adsorption energies;
    $UQ$ represents \gls{ML}-predicted uncertainty quantifications;
    $\mu$ represents the mean of a sample of points;
    $\sigma$ represents the standard deviation of a sample of points;
    $\epsilon$ represents the residuals between \gls{DFT} and \gls{ML};
    and $\hat{\epsilon}$ represents the residuals between \gls{ML}-predicted $\epsilon$ and the actual $\epsilon$.
    }\label{fig:methods}
\end{figure}

We explore various methods that aim to quantify the uncertainty for regression procedures where the predicted quantity is a continuous variable.
To standardize the assessment of performance, we ensure that each \gls{UQ} method returns predictive uncertainty results in a consistent format:  a distribution over possible outcomes of the predicted quantity for any specified input point.
This result format allows us to compute all the predictive uncertainty performance metrics which we introduce in subsequent sections.
Figure~\ref{fig:methods} illustrates all of the methods we investigate in this study, and we describe each method in detail below.

\textbf{\glsentrytext{NN}:}
To establish a baseline for predictive accuracy, we re-trained a previously reported \gls{NN}\cite{Xie2018, Back2019} on this study's training set.
This \gls{NN} model projects a three-dimensional atomic structure into a graph, which is then fed into convolutional layers to extract local atomic information for predicting global target properties.
In this case, we predict \gls{DFT}-calculated adsorption energies, $\Delta$E.
The graph consists of nodes representing atoms and edges representing distances between atoms. 
The \gls{NN} updates the node features using the local information extracted in the convolutional layers, then hidden layers in the \gls{NN} maps the node features to the adsorption energies.
Reference \citet{Back2019} for additional details.

\textbf{\glsentrytext{NN} Ensemble:}
We created an ensemble of \gls{NN}s by 5-fold subsampling the training data and then training individual \gls{NN} models on the 5 folds.
Each individual \gls{NN} model's architecture is identical to the base \gls{NN} architecture outlined previously.
The only differences are their training sets and their individually randomized initial weights.
For the final prediction of the ensemble we computed the mean of the set of models' predictions, and for the ensemble's estimate of uncertainty we computed the standard deviation of the set of predictions.

\textbf{\glsentrytext{BNN}:}
The aim of \gls{BNN} is to determine the posterior distribution of model parameters rather than a single optimal value of the parameters.
In practice, inferring true posterior distributions is very difficult and even infeasible in most cases.
Thus, we approximate the model posterior to be as close as possible to the true posterior.
The same \gls{NN} architecture was used, but we converted the \gls{NN} into \gls{BNN} by assigning posterior distributions to all model parameters in the hidden layers in the \gls{NN} model. 
The \gls{BNN} then approximated the true posterior distributions using variational inference so that it could use the approximated posterior to predict the adsorption energies.
We sampled the model parameters 20 times from the approximated posterior distributions, and used the mean of these predictions as the final prediction and the standard deviation of these predictions as the estimation of uncertainty.
We implemented the \gls{BNN} and performed variational inference using Pyro.\cite{Bingham2018}

\textbf{\glsentrytext{dropout}:}
\gls{dropout} have been shown to approximate Bayesian models.\cite{Gal2016}
We created a \gls{dropout} by first replicating the exact architecture used to create the convolutional \gls{NN} outlined previously. 
Then we enforced a random dropout rate of 30\% in the dense hidden layers that followed the convolutional layers.
The nodes were randomly dropped out during both training and prediction.
To make predictions, we sampled the \gls{dropout} 20 times.
The mean of the predictions was used as the final prediction of the \gls{dropout}, and the standard deviation of the predictions was used as the estimation of uncertainty.

\textbf{\glsentrytext{dNN}:}
Suppose we have trained a \gls{NN}.
We may aim to empirically fit an additional mapping that predicts the error of the first \gls{NN}.
Here we show \gls{dNN}, which trains a secondary \gls{NN} to predict the residuals of the initial \gls{NN}.
When training the first \gls{NN}, we hold out 10\% of the training data.
Afterwards, we use the residuals of the initial \gls{NN} on the held-out portion as training data for the second \gls{NN}.
After the secondary training, this second \gls{NN} can predict residuals for the first \gls{NN} on some new set of input data.
The predictions of the second \gls{NN} can then be used as uncertainty estimates.
Note that both the \gls{NN}s included within the \gls{dNN} were constructed using the same convolutional architecture outlined previously.

\textbf{\glsentrytext{GP}:}
\gls{GP}s are one of the most common regression methods for producing \gls{UQ}s, and so we use them here as a baseline.
We fit a standard \gls{GP} using the same exact features that we used in previous work.\cite{Tran2018}
These features are defined by the elements coordinated with the adsorbate and by the elements of its next-nearest neighbors.
Specifically:  We use the atomic numbers of these elements, their Pauling electronegativity, a count of the number of atoms of each element near the adsorbate, and the median adsorption energy between the adsorbate and the elements.
To ensure that these features interacted well with the \gls{GP}'s kernel, we normalized each of the features to have a mean of zero and standard deviation of one.
Reference \citet{Tran2018} for additional details.
To define the \gls{GP}, we assumed a constant mean and used a Matern covariance kernel.
We trained the length scale of the Matern kernel using the \gls{MLE} method.
All \gls{GP} training and predictions were done with GPU acceleration as implemented in GPyTorch.\cite{Gardner2018}.

\textbf{\glsentrytext{dGP}:}
\gls{GP}s are Bayesian models in which a prior distribution is first specified and then updated given observations to yield a posterior distribution. 
The mean of this posterior distribution is used for regression, and the covariance matrix is used for \gls{UQ}. 
Typically, in lieu of any additional prior knowledge, practitioners will take the prior distribution to have zero-mean. 
However, we could instead supply an alternative curve for the prior mean, and then perform the usual Bayesian updates to compute the posterior of this \gls{GP} given observations. 
Here, for the \gls{GP} prior mean, we supply the prediction given by a single pre-trained \gls{NN}. 
We call this method \gls{dGP}.
For the input features of this \gls{GP}, we used the same exact features we used for the plain \gls{GP}---i.e., the vector of atomic numbers, electronegativity, etc.\
For the covariance kernel of this \gls{GP}, we used a Matern kernel where we fit the kernel hyperparameters using \gls{MLE}. 
All \gls{GP} training and predictions were done with GPU acceleration as implemented in GPyTorch.\cite{Gardner2018}.

\textbf{\glsentrytext{CFGP}:}
A limitation of using this formulation of a \gls{GP} with \gls{NN}-predicted mean is that it requires the use of hand-crafted features for the \gls{GP}.
This requirement reduces the transferability of the method to other applications where such features may not be readily available.
To address this, we formulated a different method where we first train a \gls{NN} (as described previously) to predict adsorption energies and then fix the network's weights.
Then we use the 46 pooled outputs of the convolutional layers of the network as features in a new \gls{GP}.
The \gls{GP} would then be trained to use these features to produce both mean and uncertainty predictions on the adsorption energies.
We call this a \gls{CFGP}.
Note that we normalized the 46 convolution outputs of the \gls{NN} so that each output would have a mean of zero and a standard deviation of one across the training set.
To define the \gls{GP}, we assumed a constant mean and used a Matern covariance kernel.
We trained the length scale of the Matern kernel using the \gls{MLE} method.
All \gls{GP} training and predictions were done with GPU acceleration as implemented in GPyTorch.\cite{Gardner2018}.

\subsection{Performance metrics}

We used five different metrics to quantify the accuracy of the various models:  \gls{MDAE}, \gls{RMSE}, \gls{MAE}, \gls{MARPD}, and \gls{R2}.
We used \gls{MDAE} because is insensitive to outliers and is therefore a good measure of accuracy for the majority of the data.
We used \gls{RMSE} because it is sensitive to outliers and is therefore a good measure of worst-case accuracy.
We used \gls{MAE} because it lies between \gls{MDAE} and \gls{RMSE} in terms of sensitivity to outliers.
We used \gls{MARPD} and \gls{R2} because they provide normalized measures of accuracy that may be more interpretable for those unfamiliar with adsorption energy measurements in \gls{eV}.
\gls{MARPD} values were calculated with Equation~\ref{eq:marpd}: 

\begin{equation}\label{eq:marpd}
    MARPD = \frac{1}{N} \sum_{n=1}^{N} \abs{100 \cdot \frac{\hat{x}_n - x_n}{\abs{\hat{x}_n} + \abs{x_n}}}
\end{equation}

\noindent where $n$ is the index of a data point, $N$ is the total number of data points, $x_n$ is the true value of the data point, and $\hat{x}_n$ is the model's estimate of $x_n$.
In this case, $x_n$ is a DFT-calculated adsorption energy and $\hat{x}_n$ is the surrogate-model-calculated adsorption energy.
The ensemble of these metrics provide a more robust view of accuracy than any one metric can provide alone.

To assess the calibration (or ``honesty'') of these models' \gls{UQ}s, we created calibration curves.
A calibration curve ``displays the true frequency of points in each interval relative to the predicted fraction of points in that interval'', as outlined by \citet{Kuleshov2018}.
In other words:  We used the standard deviation predictions to create Gaussian random variables for each test point and then tested how well the residuals followed their respective Gaussian random variables.
Thus ``well-calibrated'' models had residuals that created Gaussian distributions whose standard deviations were close to the model's predicted standard deviations.
We discuss calibration curves in more detail in the Results section alongside specific examples.
We also calculated the calibration errors\cite{Kuleshov2018} of our models, which is a quantitative measure of calibration.

As \citet{Kuleshov2018} also pointed out, well-calibrated models are necessary but not sufficient for useful \gls{UQ}s.
For example:  A well-calibrated model could still have large uncertainty estimates, which are inherently less useful than well-calibrated and small uncertainty estimates.
This idea of having small uncertainty estimates is called ``sharpness'', and \citet{Kuleshov2018} define it with Equation~\ref{eq:og_sharpness}:

\begin{equation}\label{eq:og_sharpness}
    sha = \frac{1}{N} \sum_{n=1}^{N} var(F_n)
\end{equation}

\noindent where $var(F_n)$ is the variance of the random variable whose cumulative distribution function is $F$ at point $n$.
This is akin to the average variance of the uncertainty estimates on the test set.
Here we propose and use a new formulation (Equation~\ref{eq:sharpness}) where we add a square root operation.
This operation gives the sharpness the same units as the predictions, which provides us with a more intuitive reference.
In other words:  Sharpness is akin to the average of the \gls{ML}-predicted standard deviations.

\begin{equation}\label{eq:sharpness}
    sha = \sqrt{\frac{1}{N} \sum_{n=1}^{N} var(F_n)}
\end{equation}

Another consideration is the dispersion of the uncertainty estimates.
If a model predicts a constant value for uncertainty, it may still be able to perform well with regards to calibration or sharpness.
Constant values for uncertainty are likely to fail when models are used to make predictions outside the bounds of the training data.
One way to address this issue is to calculate the \gls{Cv}\cite{Levi2020}.
See Equation~\ref{eq:dispersion}:

\begin{equation}\label{eq:dispersion}
    C_v = \frac{\sqrt{\frac{\sum_{n=1}^{N}(\sigma_n-\mu_{\sigma})^2}{N-1}}}{\mu_{\sigma}}  
\end{equation}

\noindent where $\sigma_n$ is the predicted standard deviation of point $n$, $\mu_{\sigma}$ is the average value of $\sigma_n$, and $N$ is the total number of test points.
Low values of \gls{Cv} indicate a narrow dispersion of uncertainty estimates, which may suggest poor performance in out-of-domain predictions.
Thus a higher \gls{Cv} may indicate more robust uncertainty estimates.
But as \citet{Scalia2019} point out, the optimal dispersion is a function of the validation/test data distribution.
Therefore, \gls{Cv} should be used as a secondary screening metric rather than a primary performance metric.

We also assessed the performance of each predictive uncertainty method by comparing their \gls{NLL} values the test set.
For each test point, we established a Gaussian probability distribution using the mean and uncertainty predictions of each \gls{UQ} model.
Then we calculated the conditional probability of observing the true value of the test point given the probability distribution created from the \gls{UQ}; this is the likelihood of one test point.
We then calculated the product of all the likelihoods of all test points, which yielded the total test likelihood.
It follows that better \gls{UQ} methods yield higher total likelihood values.
Equivalently, we could calculate the natural logarithms of each likelihood, sum them, and then take the negative of this value; this is \gls{NLL}.
Equation~\ref{eq:nll} shows how we calculated \gls{NLL}:

\begin{equation}\label{eq:nll}
    NLL = - \sum_{i=1}^{n} \ln{P(y_i | N(\hat{y_i}, \hat{\sigma_i}^2))}
\end{equation}

\noindent where $y_i$ is the true value of a test point, $\hat{y}_i$ is a model's predicted mean value at that test point, $\hat{\sigma_i}^2$ is the model's predicted variance at that test point, $n$ is the set of all test points, and $N(x, y)$ is a normal distribution with mean $x$ and variance $y$.
Note how the \gls{NLL} value depends on the size and location of the test set.
This means that the absolute value of \gls{NLL} changes from application to application, and so a ``good'' \gls{NLL} value must be contextualized within a particular test set.
Within a test set, a lower \gls{NLL} value indicates a better fit.
We also note that we assumed Gaussian distributions for our \gls{UQ} methods' predictions.
This assumption does not necessarily need to be applied, meaning that the normal distribution in Equation~\ref{eq:nll} may be replaced with any other appropriate distribution.

We use \gls{NLL} because it provides an overall assessment that is influenced by both the predictive accuracy of a method as well as the quality of its \gls{UQ}.
Previous work~\cite{Gneiting2007, Dawid2014} has shown the \gls{NLL} to be a strictly proper scoring rule, which intuitively means that it provides a fair quantitative assessment (or score) for the performance of the \gls{UQ} method, and that it can be decomposed into terms that relate to both calibration and sharpness.
\gls{NLL} is also a popular performance metric that has been used to quantify uncertainty in a variety of prior work~\cite{Lakshminarayanan2017} and provides an additional single score for \gls{UQ} methods.


\section{Results}

\subsection{Illustrative examples}

\begin{figure}
    \centering
    \begin{subfigure}[b]{0.32\textwidth}
        \includegraphics[width=\textwidth]{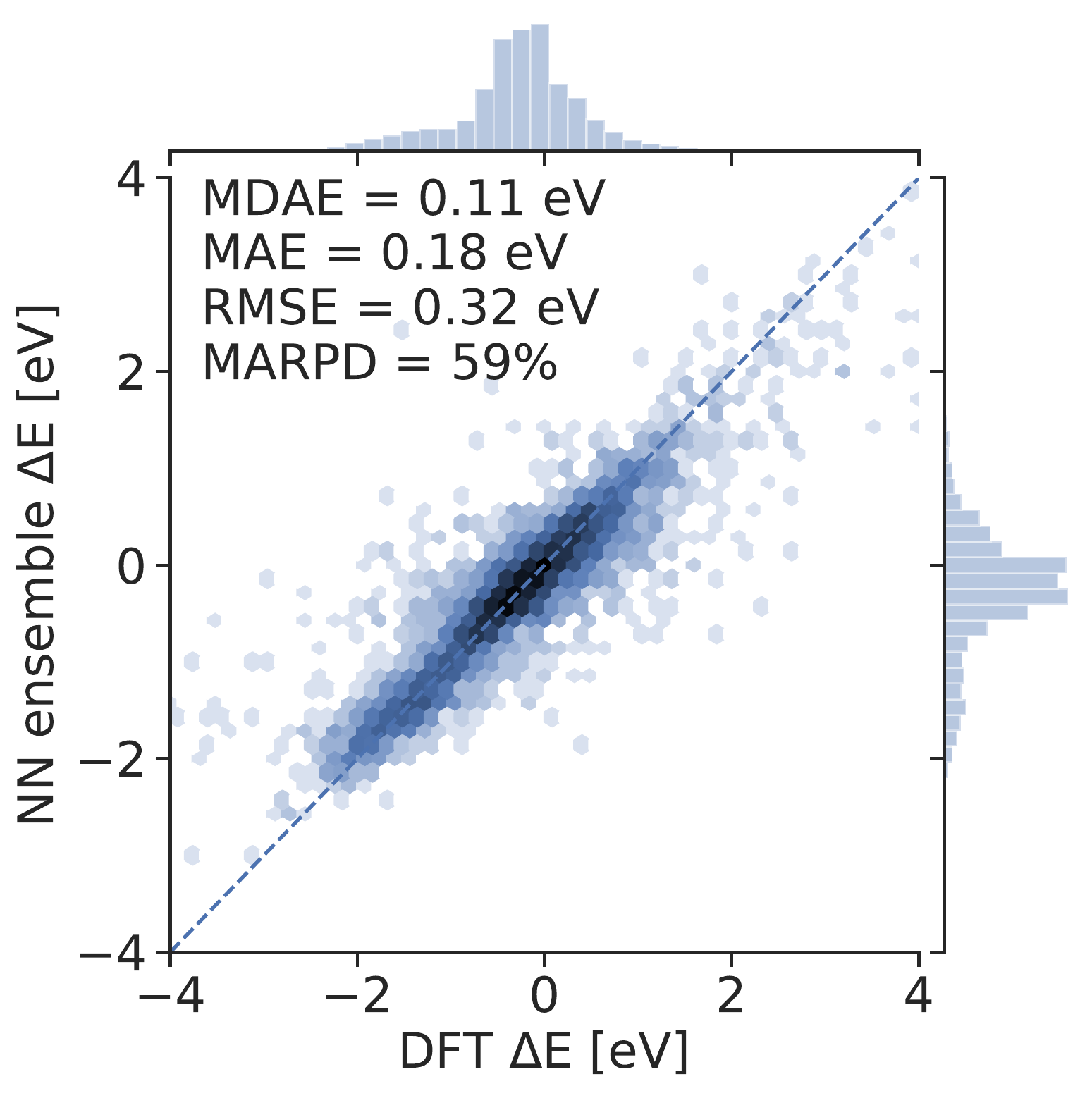}
        \caption{parity plot}\label{fig:parity_example}
    \end{subfigure}
    \begin{subfigure}[b]{0.32\textwidth}
        \includegraphics[width=\textwidth]{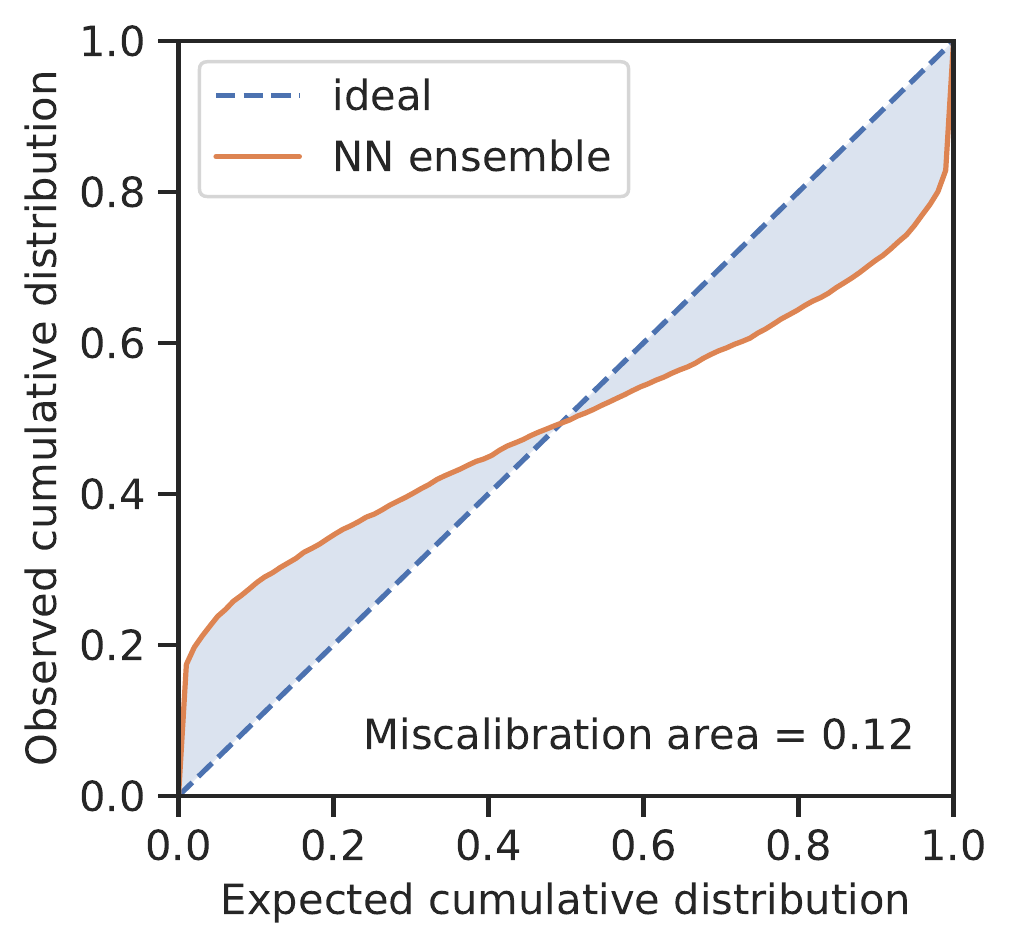}
        \caption{calibration curve}\label{fig:calibration_example}
    \end{subfigure}
    \begin{subfigure}[b]{0.32\textwidth}
        \includegraphics[width=\textwidth]{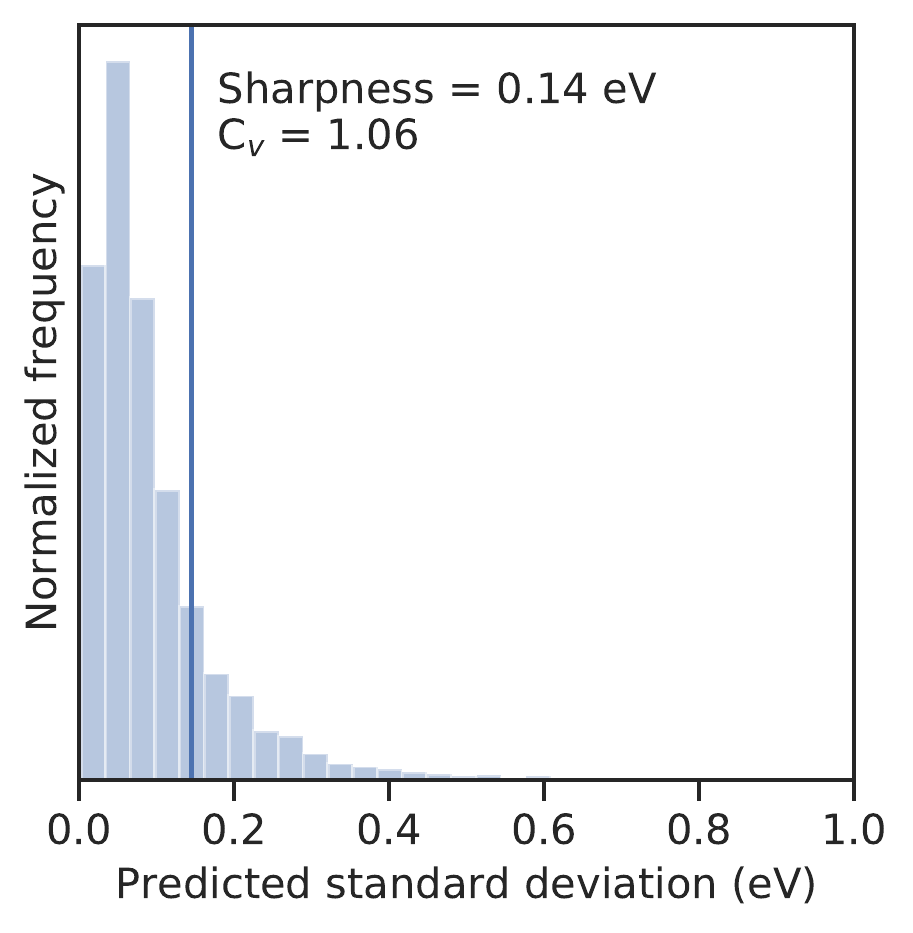}
        \caption{uncertainty distribution}\label{fig:sharpness_example}
    \end{subfigure}
    \caption{Results of the \gls{NN} ensemble. Each figure here was created with the test set of 8,289 points.}\label{fig:results_example}
\end{figure}

Let us first discuss the results of our \gls{NN} ensemble for illustrative purposes.
Figure~\ref{fig:results_example} contains a parity plot, calibration curve, and predicted-uncertainty distribution of our \gls{NN} ensemble model.
The parity plot shows the accuracy of the model; the calibration curve shows the honesty of the model's uncertainty predictions; and the uncertainty distribution shows the sharpness of the model's uncertainty predictions.
Accurate models have parity plots whose points tend to fall near the diagonal parity line.
Calibrated models have calibration curves that approach the ideal diagonal line.
Sharp models have uncertainty distributions that tend towards zero.
Note that sharpness should not be won at the cost of calibration.

The calibration curve was created by first establishing Gaussian random variables for each test point where the means were the model's predictions and the variances were the model's predicted variances.
The test residuals could then be compared against their respective random variables.
For simplification purposes, we divided each of the test residuals by their corresponding standard deviations so that we could test all residuals against the same unit Gaussian distribution.
Thus if the normalized test residuals followed a unit Gaussian distribution, then the model's uncertainty predictions could be considered well-calibrated.
We tested this by calculating the theoretical cumulative distribution of points within the intervals $(-\infty, x] \mkern9mu \forall \mkern9mu x \in (-\infty, \infty)$ and then compared it against the observed cumulative distributions.  
A plot of the observed cumulative distributions against the theoretical cumulative distributions is called a calibration curve.
A perfectly calibrated model would have normalized residuals that are Gaussian, which would yield a diagonal calibration line.
Therefore, models' calibration could be qualified by the closeness of their calibration curves to this ideal, diagonal curve.
We quantified this closeness by calculating the area between the calibration curve and the ideal diagonal.
We call this the miscalibration area, and smaller values indicate better calibration.
We also calculated the calibration error,\cite{Kuleshov2018} which is the mean square difference between the expected cumulative distributions and observed cumulative distributions.

\begin{figure}
    \centering
    \begin{subfigure}[b]{0.32\textwidth}
        \includegraphics[width=\textwidth]{calibration_ensemble.pdf}
        \caption{calibration of \gls{NN} ensemble}\label{fig:calibration_example_ensemble}
    \end{subfigure}
    \begin{subfigure}[b]{0.32\textwidth}
        \includegraphics[width=\textwidth]{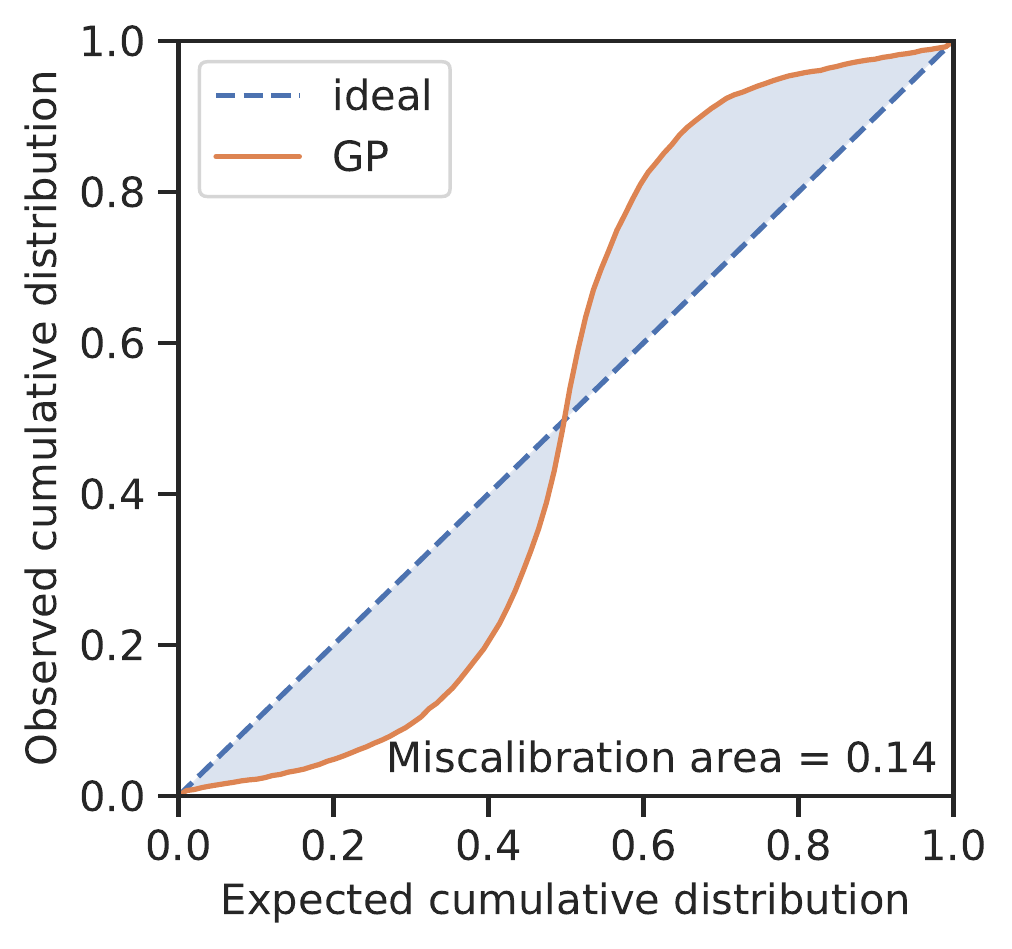}
        \caption{calibration of \gls{GP}}\label{fig:calibration_example_gp}
    \end{subfigure}
    \begin{subfigure}[b]{0.32\textwidth}
        \includegraphics[width=\textwidth]{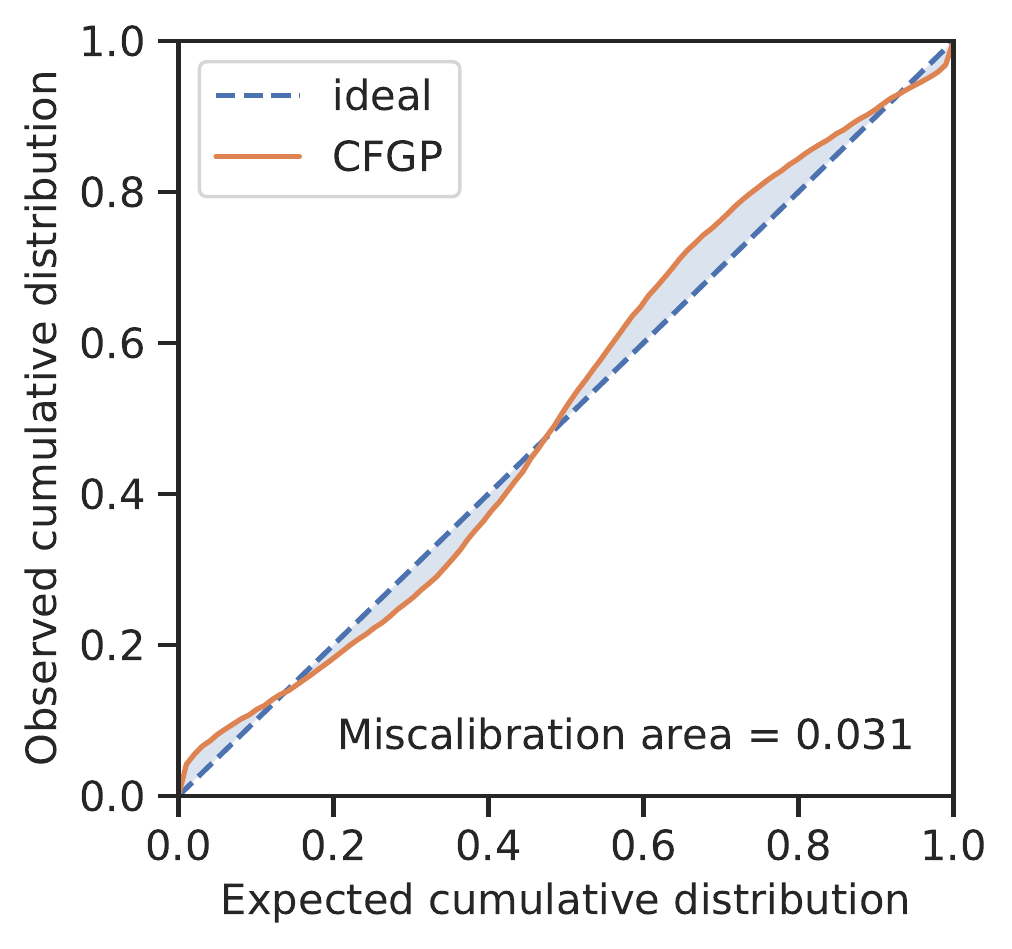}
        \caption{calibration of \gls{CFGP}}\label{fig:calibration_example_cfgp}
    \end{subfigure}
    \begin{subfigure}[b]{0.32\textwidth}
        \includegraphics[width=\textwidth]{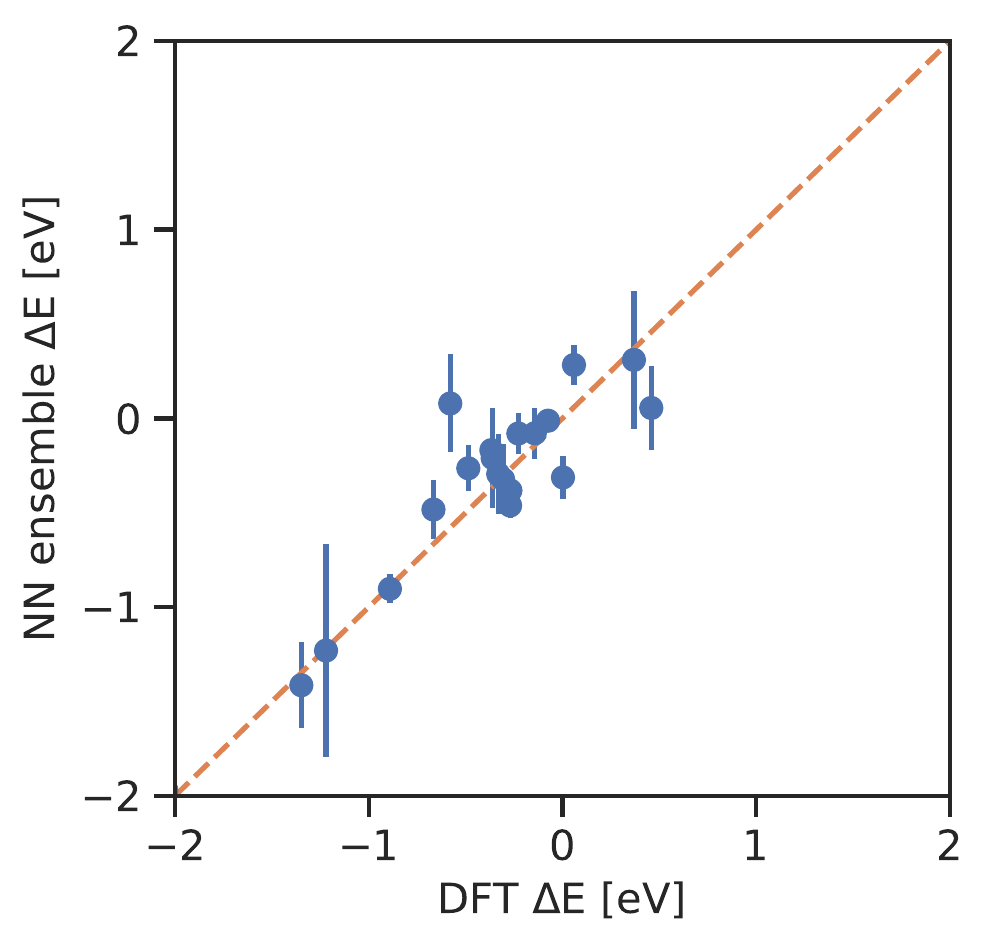}
        \caption{parity of \gls{NN} ensemble}\label{fig:error_bar_ensemble}
    \end{subfigure}
    \begin{subfigure}[b]{0.32\textwidth}
        \includegraphics[width=\textwidth]{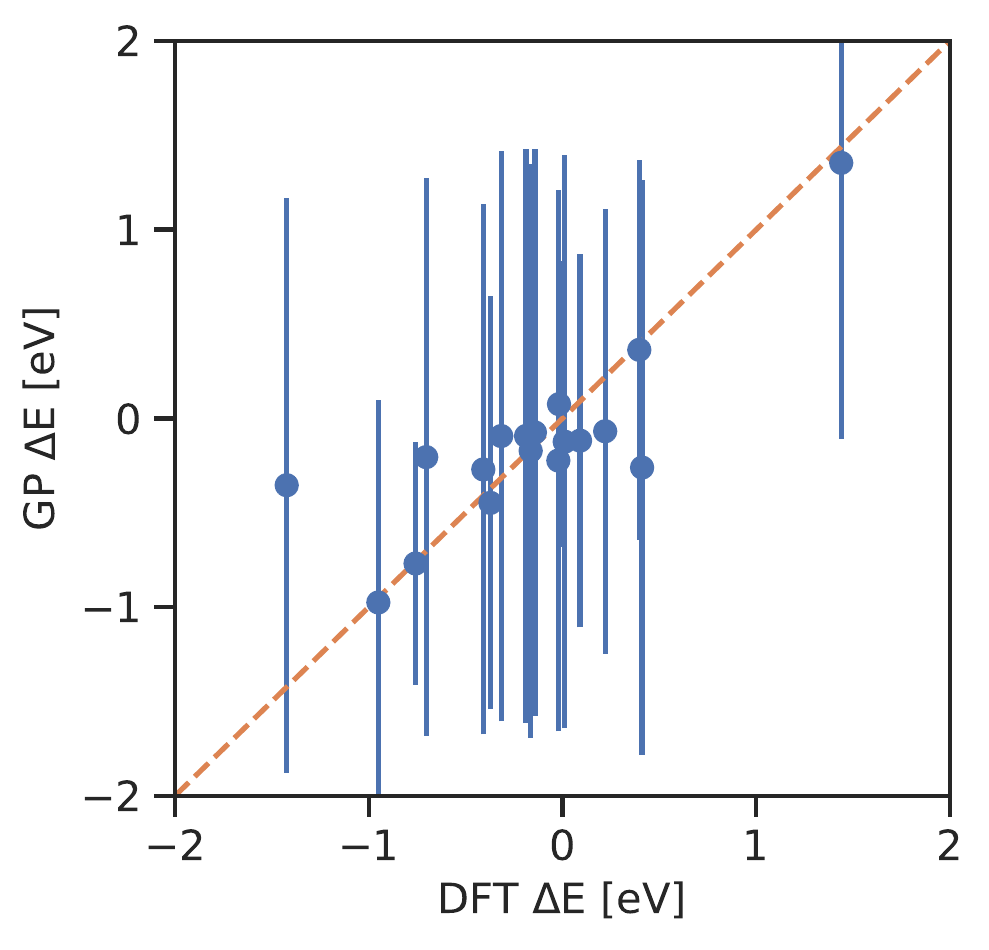}
        \caption{parity of \gls{GP}}\label{fig:error_bar_gp}
    \end{subfigure}
    \begin{subfigure}[b]{0.32\textwidth}
        \includegraphics[width=\textwidth]{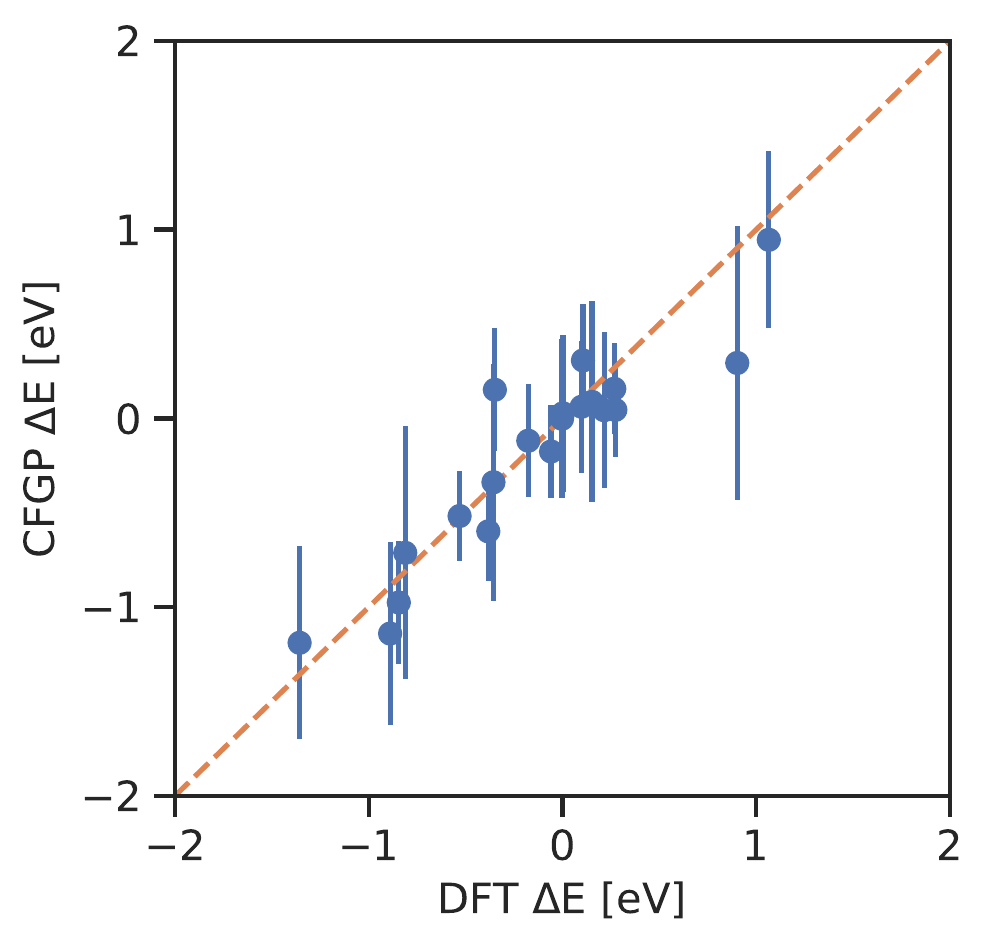}
        \caption{parity of \gls{CFGP}}\label{fig:error_bar_cfgp}
    \end{subfigure}
    \caption{Calibration curves and parity plots of an overconfident \gls{NN} ensemble, an underconfident \gls{GP}, and better-calibrated \gls{CFGP}.
    The vertical uncertainty bands in the parity plots indicate $\pm$2 standard deviations in the uncertainty predictions of each model.
    For clarity, we sampled only 20 points of the 8,289 test points to put in the parity plots.
    It follows that relatively overconfident models would have more points with uncertainty bands that do not cross the diagonal parity line;
    relatively underconfident models would have more points that cross the diagonal parity line;
    and a well-calibrated model would have \textit{ca.} 19 out of 20 points cross the parity line.}\label{fig:error_bars}
\end{figure}

The shape of a calibration curve could also yield other insights.
If a model's \gls{UQ}s were too low/confident, then the normalized residuals would be too large and they would fall outside their distributions too frequently.
This would result in a lower observed cumulative distributions compared to the expected cumulative distributions, which would correspond to a calibration curve that falls below the ideal diagonal.
Therefore, overconfident models yield calibration curves that fall under the ideal diagonal, and underconfident models yield calibration curves that fall over the ideal diagonal.
Figure~\ref{fig:error_bars} illustrates this point by plotting calibration curves of various models alongside their parity plots that contain error bars corresponding to $\pm$2 standard deviations.
Note that when we say a calibration curve ``falls under the diagonal'', we allude to curves whose right-hand-side fall under the diagonal.

\subsection{Summary results}

\begin{figure}
    \centering
    \begin{subfigure}{0.32\textwidth}
        \includegraphics[width=\textwidth]{parity_ensemble.pdf}
        \caption{\gls{NN} ensemble}\label{fig:parity_ensemble}
    \end{subfigure}
    \begin{subfigure}{0.32\textwidth}
        \includegraphics[width=\textwidth]{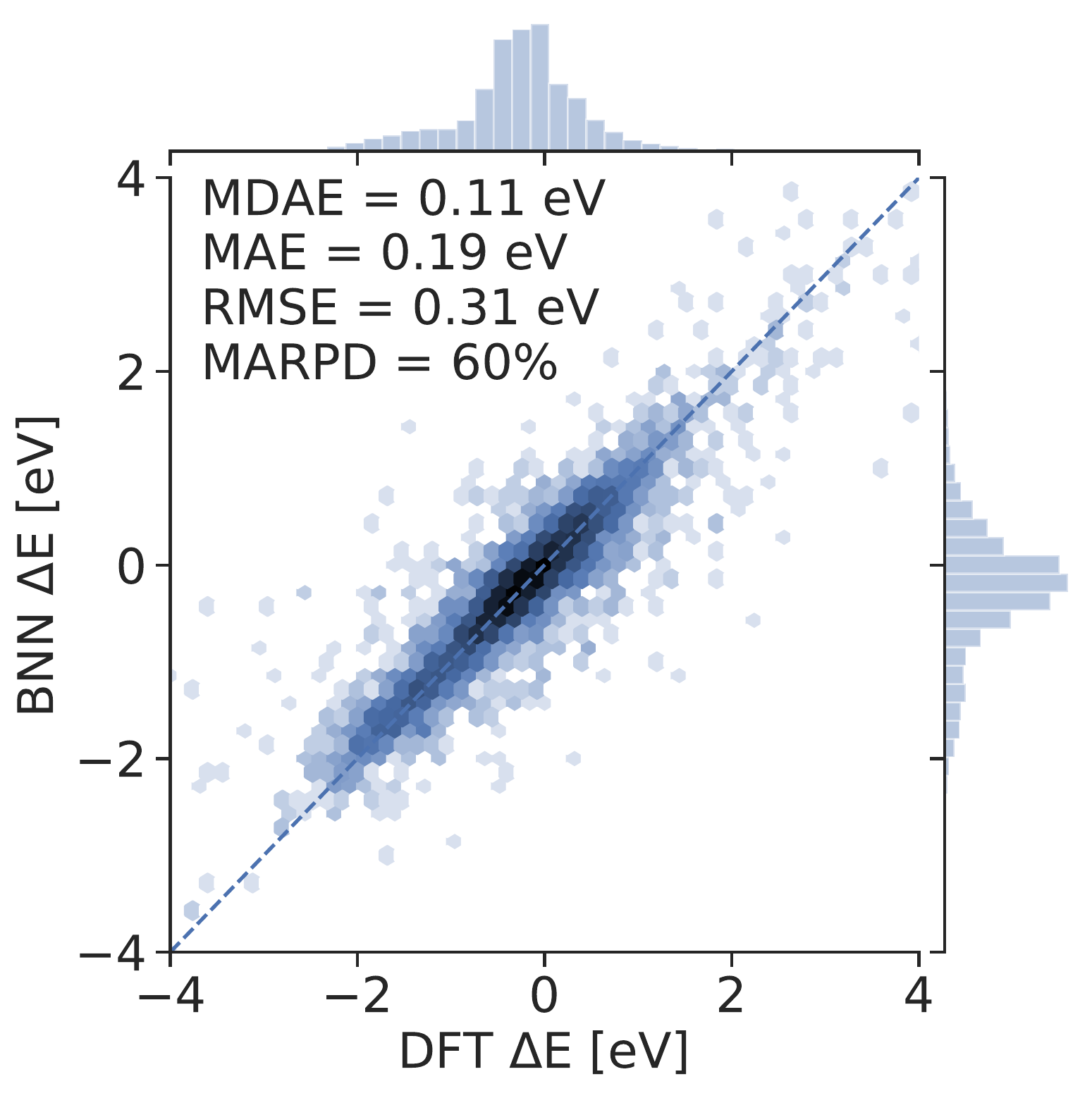}
        \caption{\gls{BNN}}\label{fig:parity_bnn}
    \end{subfigure}
    \begin{subfigure}{0.32\textwidth}
        \includegraphics[width=\textwidth]{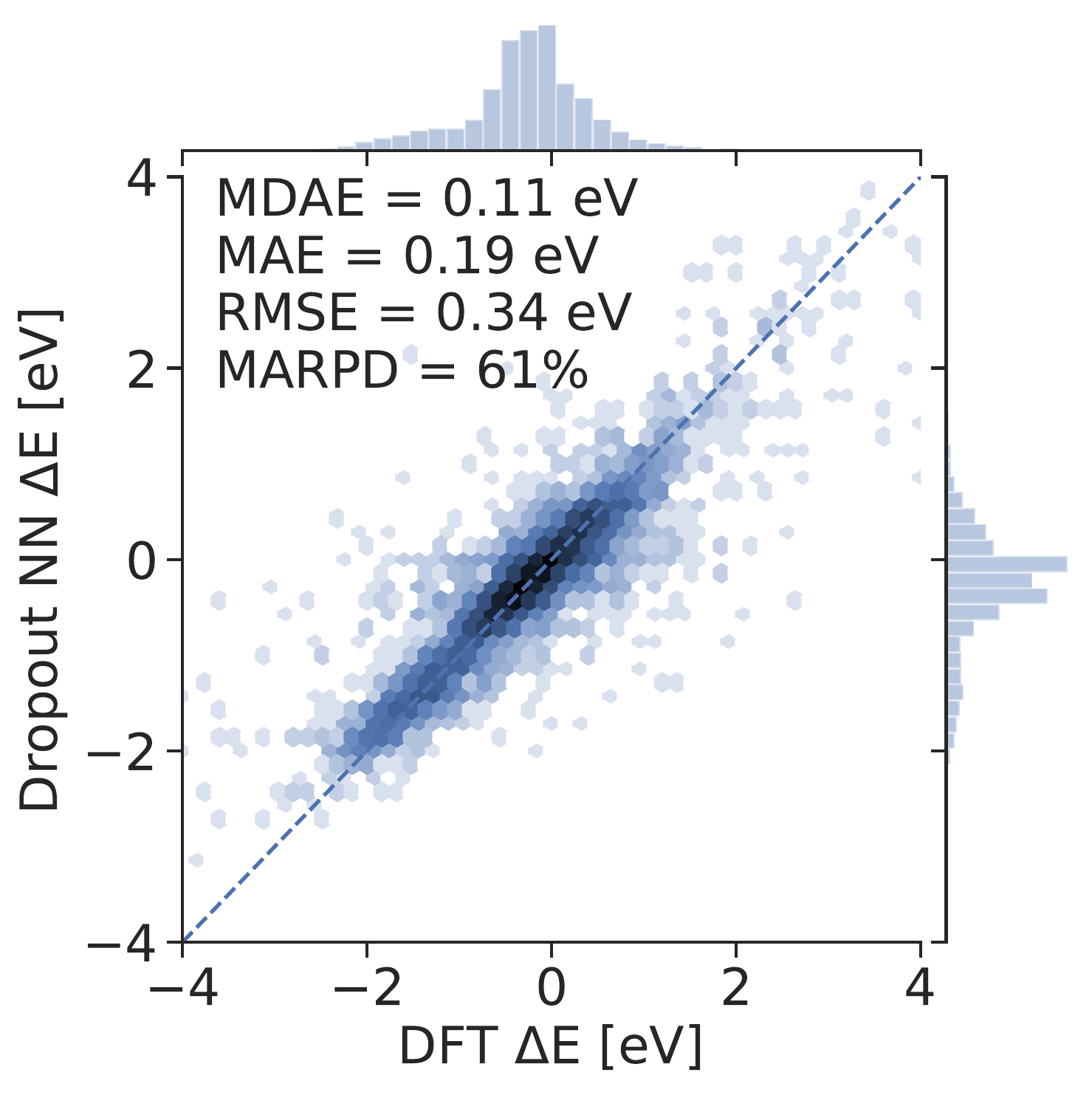}
        \caption{\gls{dropout}}\label{fig:parity_dropout}
    \end{subfigure}
    \begin{subfigure}{0.32\textwidth}
        \includegraphics[width=\textwidth]{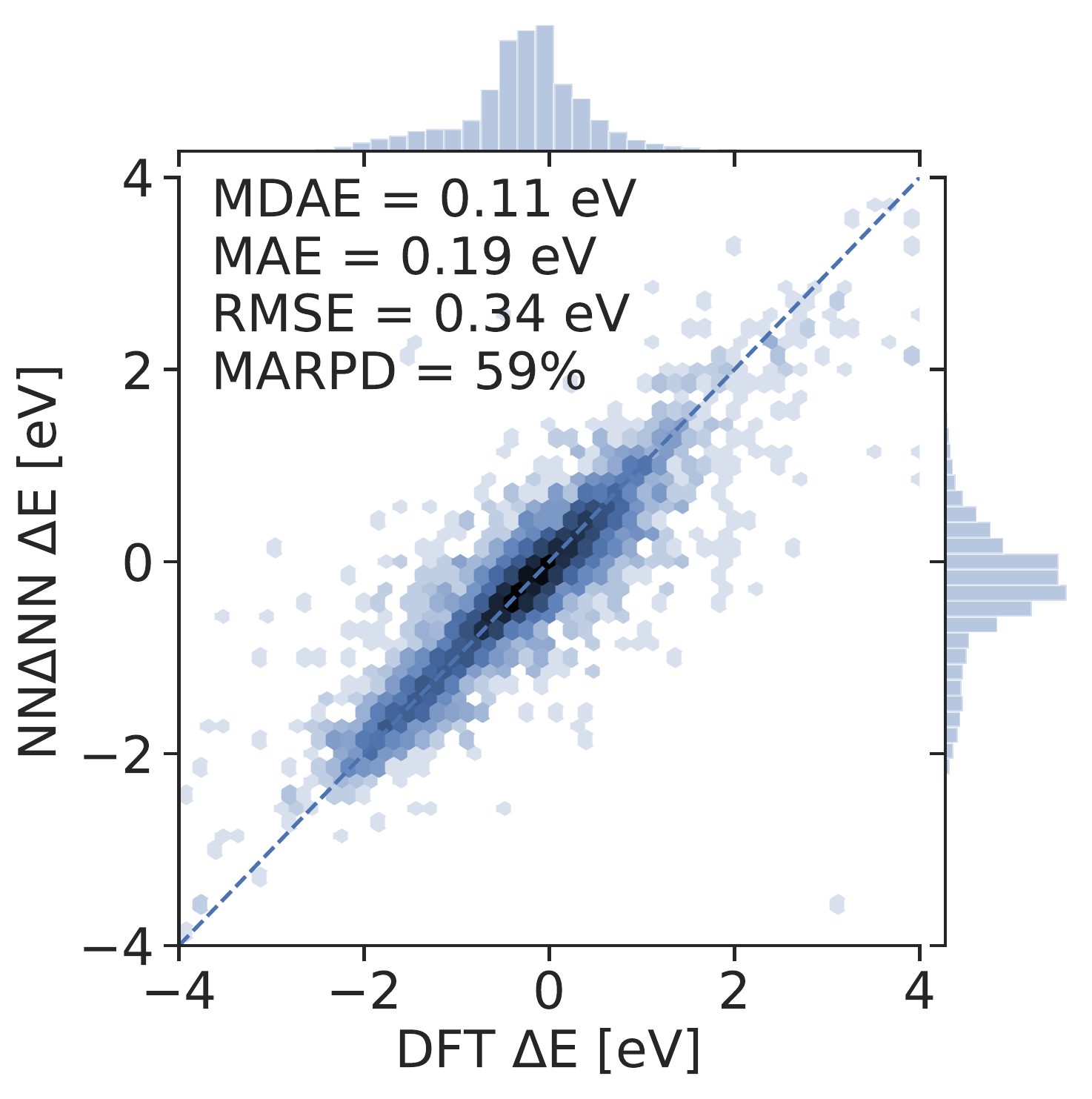}
        \caption{\gls{dNN}}\label{fig:parity_dnn}
    \end{subfigure}
    \begin{subfigure}{0.32\textwidth}
        \includegraphics[width=\textwidth]{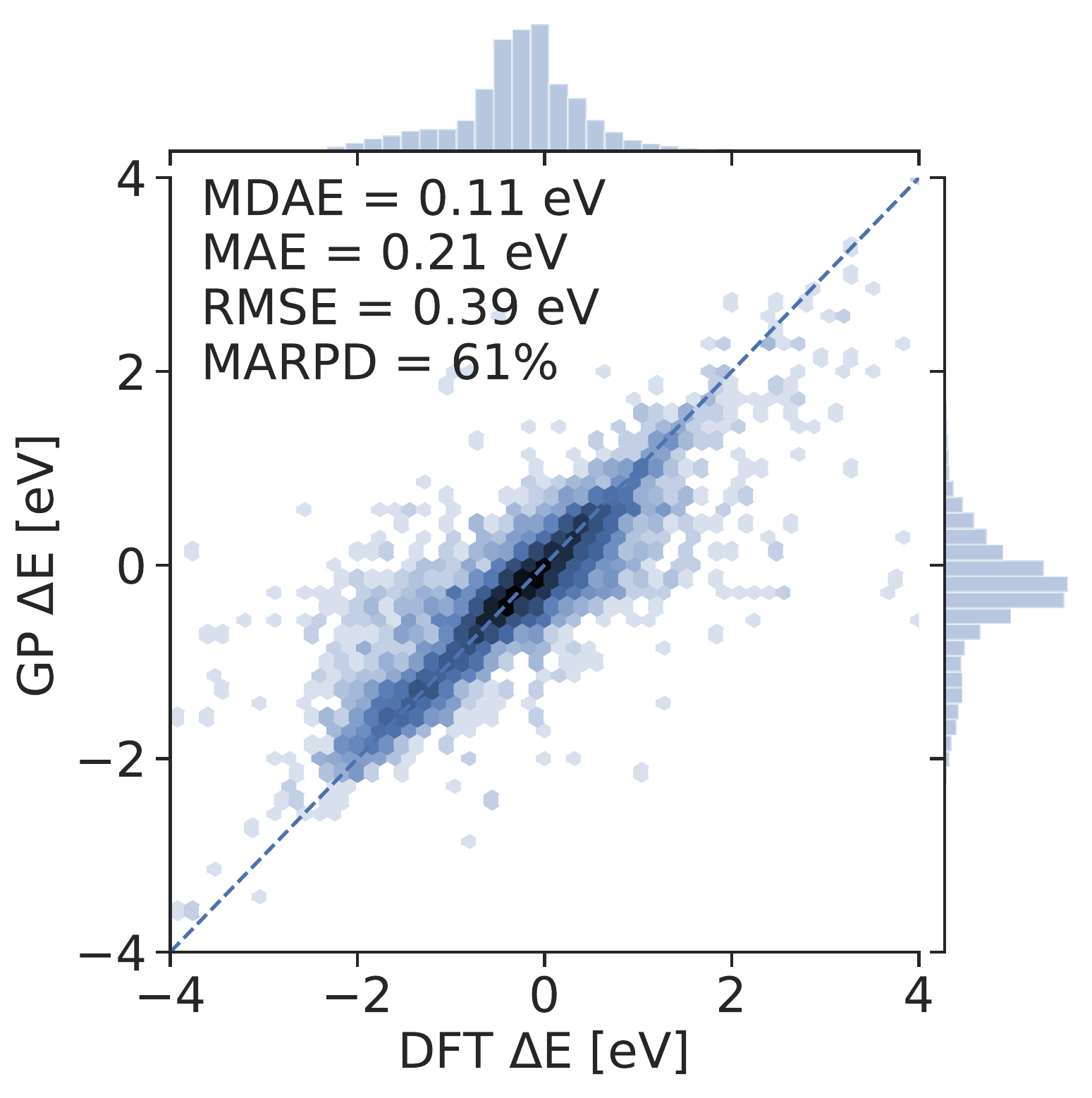}
        \caption{\gls{GP}}\label{fig:parity_gp}
    \end{subfigure}
    \begin{subfigure}{0.32\textwidth}
        \includegraphics[width=\textwidth]{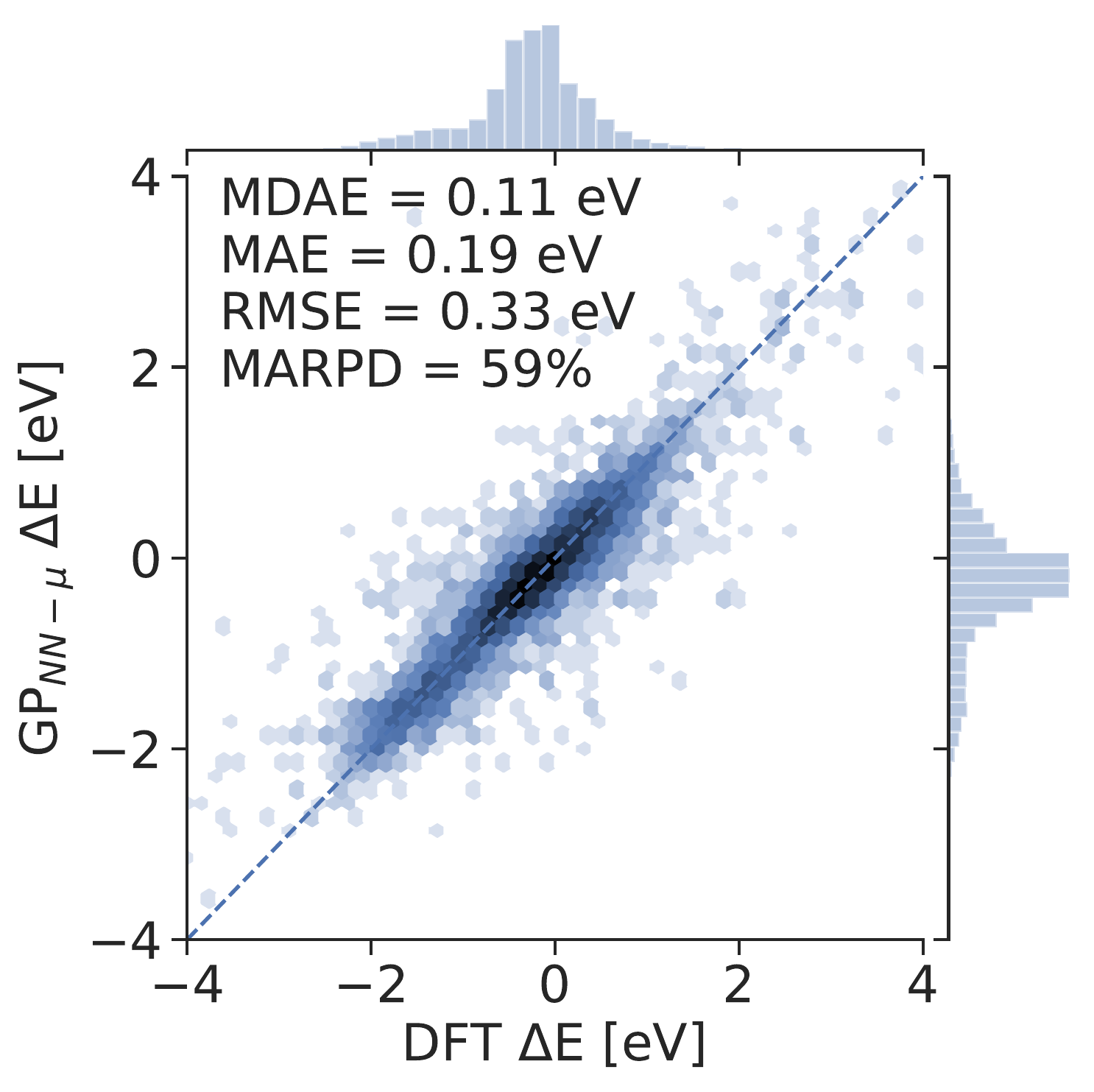}
        \caption{\gls{dGP}}\label{fig:parity_dgp}
    \end{subfigure}
    \begin{subfigure}{0.32\textwidth}
        \includegraphics[width=\textwidth]{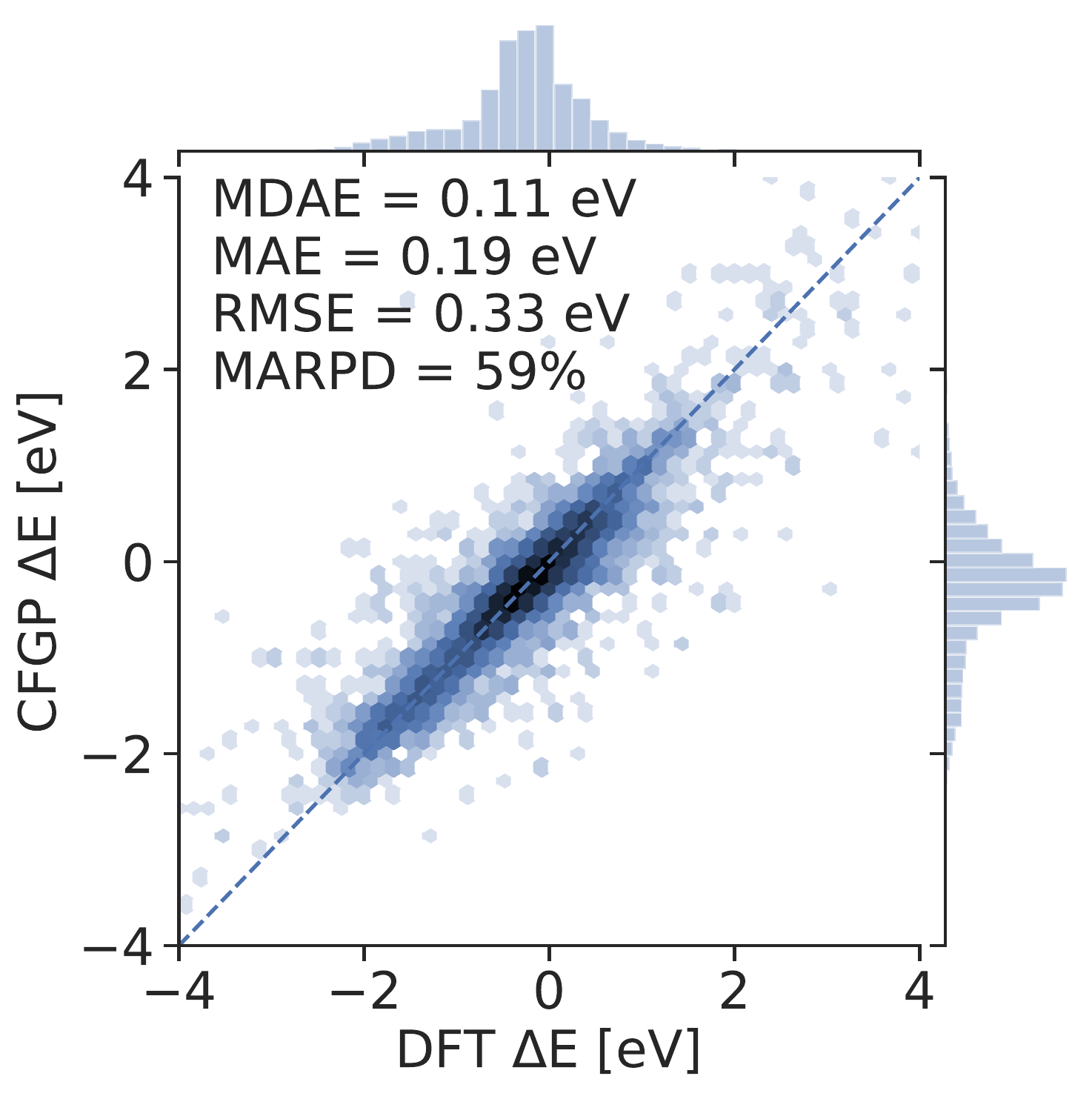}
        \caption{\gls{CFGP}}\label{fig:parity_cfgp}
    \end{subfigure}
    \caption{Parity plots for all \gls{UQ} methods used in this study.
    Shading plots were used in lieu of scatter plots because the large number of test points (8,289) obfuscated patterns.
    Darker shading indicates a higher density of points.
    Logarithmically scaled shading was used to accentuate outliers.
    The dashed, diagonal lines indicate parity.}\label{fig:parity}
\end{figure}

\begin{figure}
    \centering
    \begin{subfigure}{0.32\textwidth}
        \includegraphics[width=\textwidth]{calibration_ensemble.pdf}
        \caption{\gls{NN} ensemble}\label{fig:calibration_ensemble}
    \end{subfigure}
    \begin{subfigure}{0.32\textwidth}
        \includegraphics[width=\textwidth]{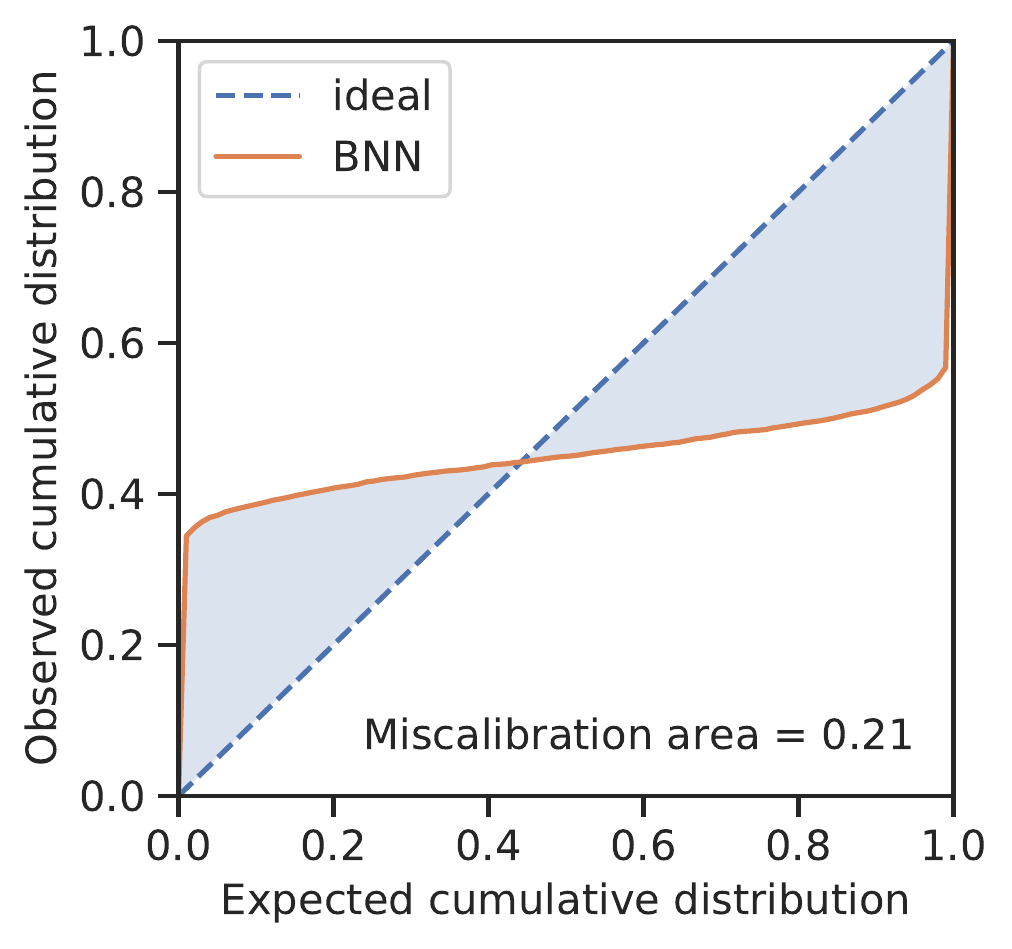}
        \caption{\gls{BNN}}\label{fig:calibration_bnn}
    \end{subfigure}
    \begin{subfigure}{0.32\textwidth}
        \includegraphics[width=\textwidth]{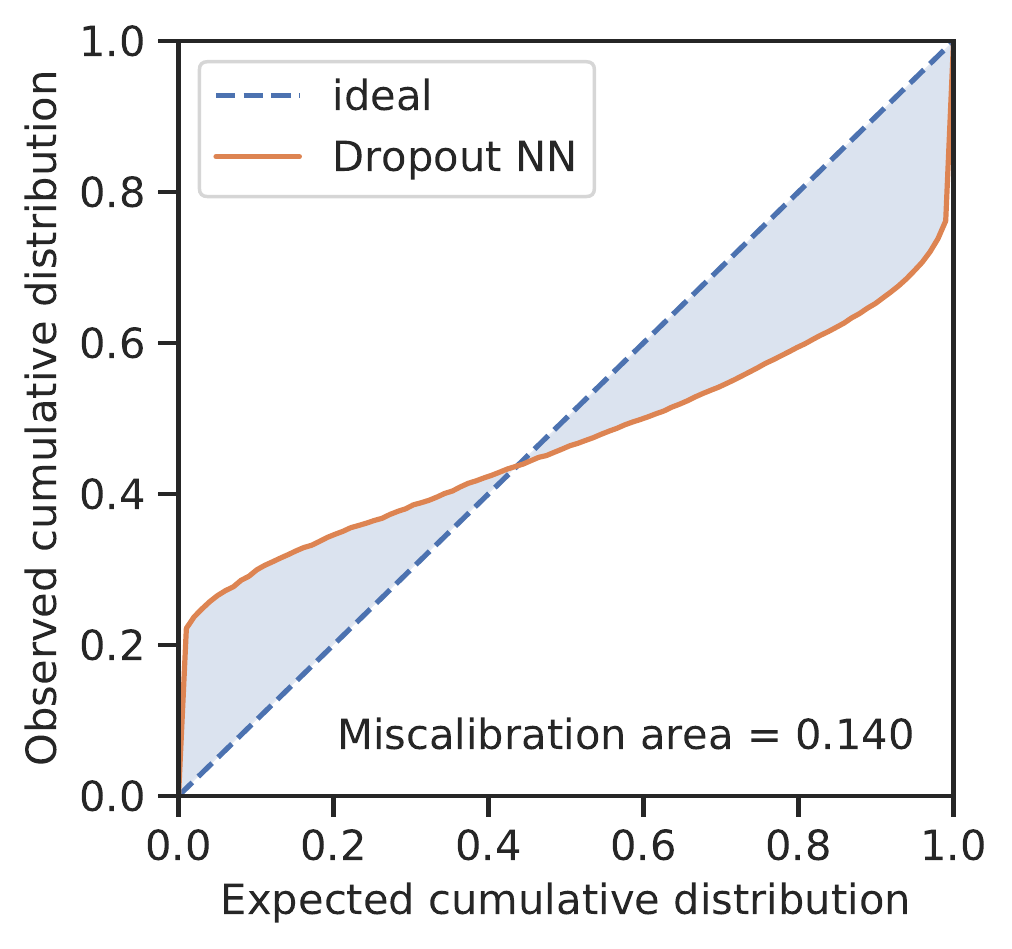}
        \caption{\gls{dropout}}\label{fig:calibration_dropout}
    \end{subfigure}
    \begin{subfigure}{0.32\textwidth}
        \includegraphics[width=\textwidth]{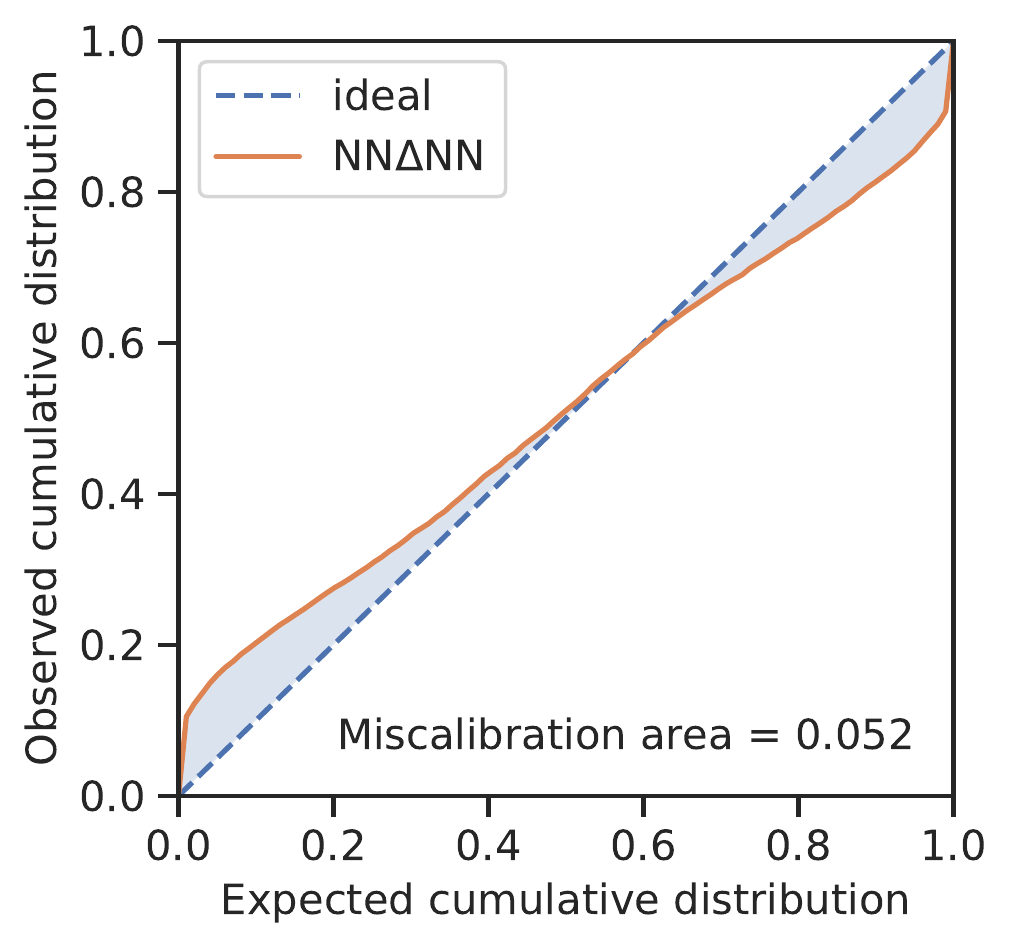}
        \caption{\gls{dNN}}\label{fig:calibration_dnn}
    \end{subfigure}
    \begin{subfigure}{0.32\textwidth}
        \includegraphics[width=\textwidth]{calibration_gp.pdf}
        \caption{\gls{GP}}\label{fig:calibration_gp}
    \end{subfigure}
    \begin{subfigure}{0.32\textwidth}
        \includegraphics[width=\textwidth]{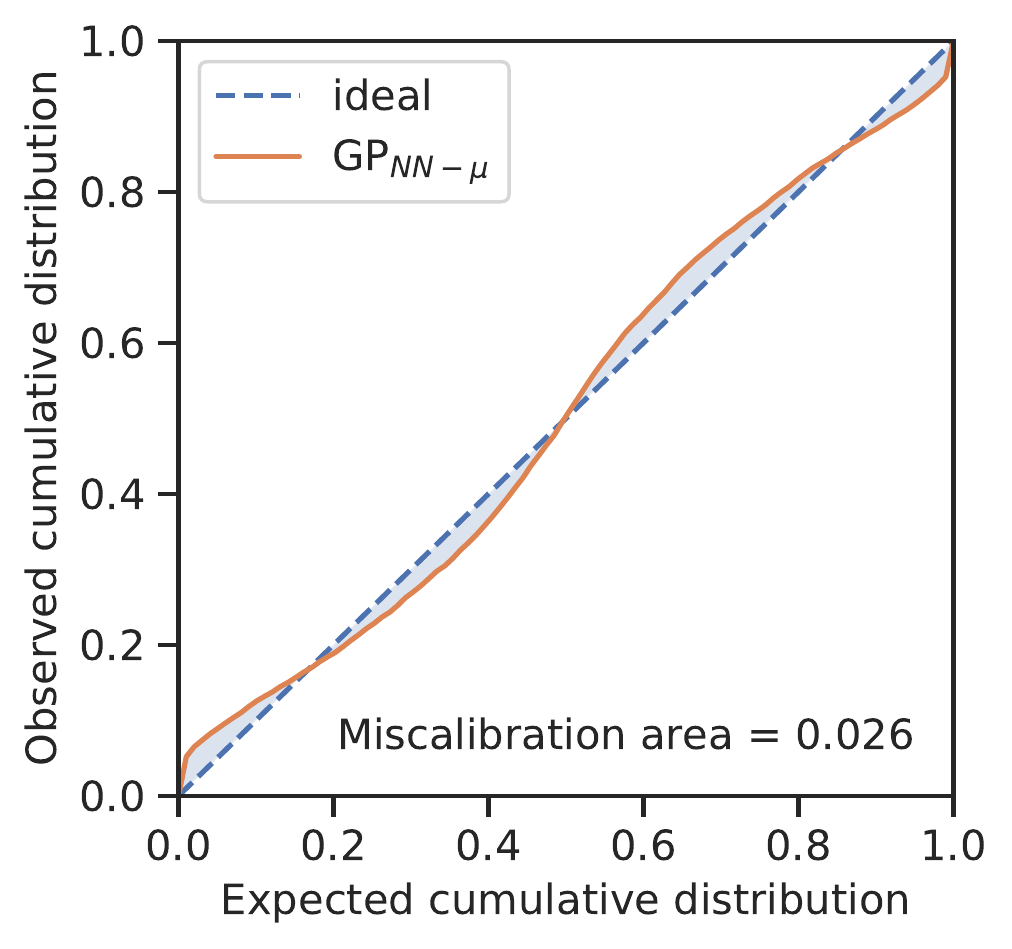}
        \caption{\gls{dGP}}\label{fig:calibration_dgp}
    \end{subfigure}
    \begin{subfigure}{0.32\textwidth}
        \includegraphics[width=\textwidth]{calibration_cfgp.pdf}
        \caption{\gls{CFGP}}\label{fig:calibration_cfgp}
    \end{subfigure}
    \caption{Calibration curves for all \gls{UQ} methods used in this study.
    Dashed, blue lines indicate perfect calibration while solid orange lines indicate the experimental calibration of the test set.
    The blue, shaded area between these lines is defined as the miscalibration area.
    }\label{fig:calibration}
\end{figure}

\begin{figure}
    \centering
    \begin{subfigure}{0.32\textwidth}
        \includegraphics[width=\textwidth]{sharpness_ensemble.pdf}
        \caption{\gls{NN} ensemble}\label{fig:sharpness_ensemble}
    \end{subfigure}
    \begin{subfigure}{0.32\textwidth}
        \includegraphics[width=\textwidth]{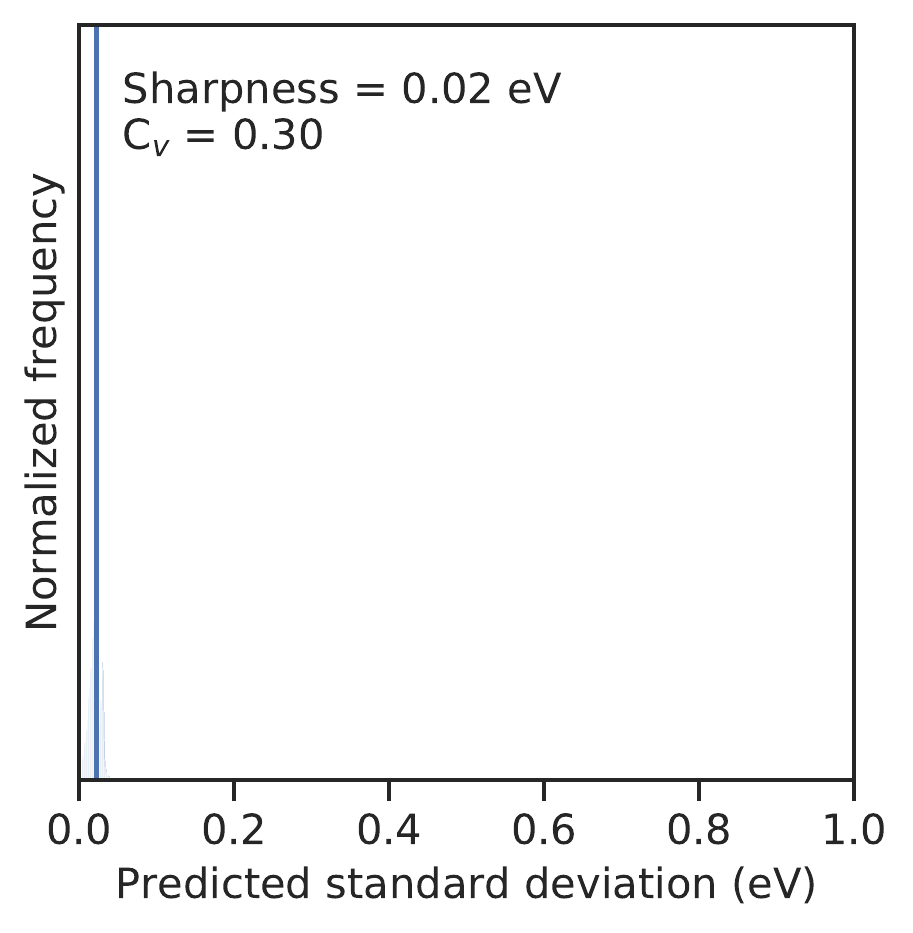}
        \caption{\gls{BNN}}\label{fig:sharpness_bnn}
    \end{subfigure}
    \begin{subfigure}{0.32\textwidth}
        \includegraphics[width=\textwidth]{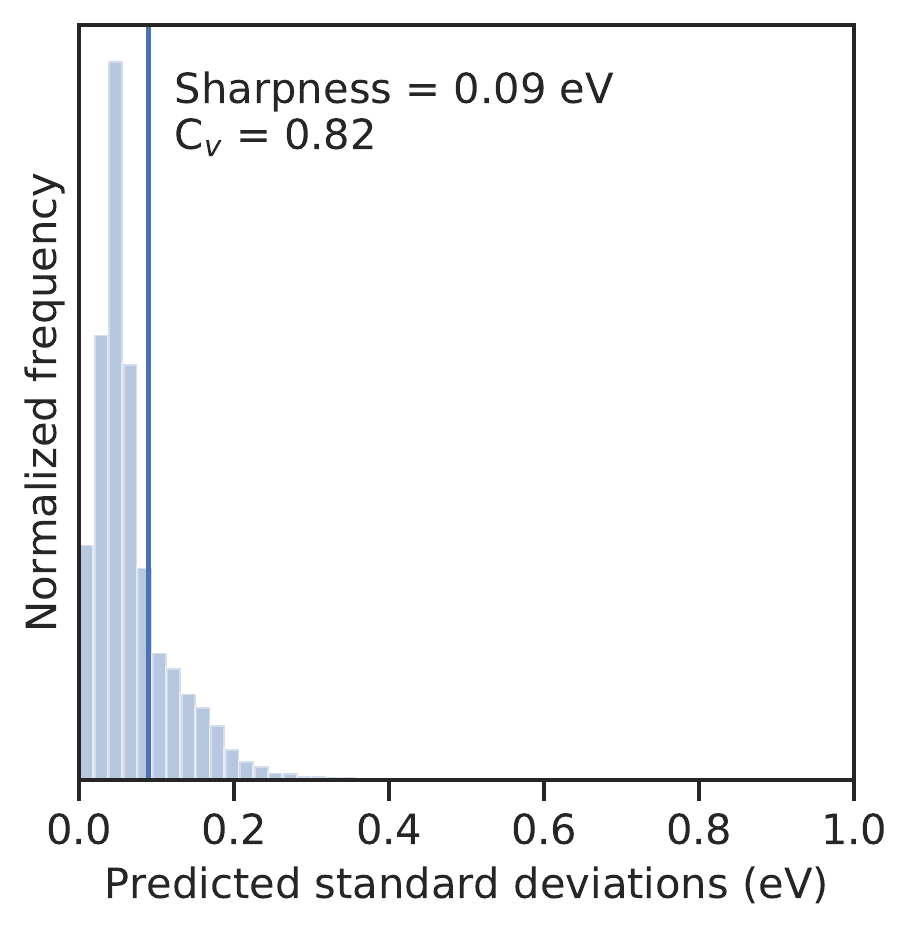}
        \caption{\gls{dropout}}\label{fig:sharpness_dropout}
    \end{subfigure}
    \begin{subfigure}{0.32\textwidth}
        \includegraphics[width=\textwidth]{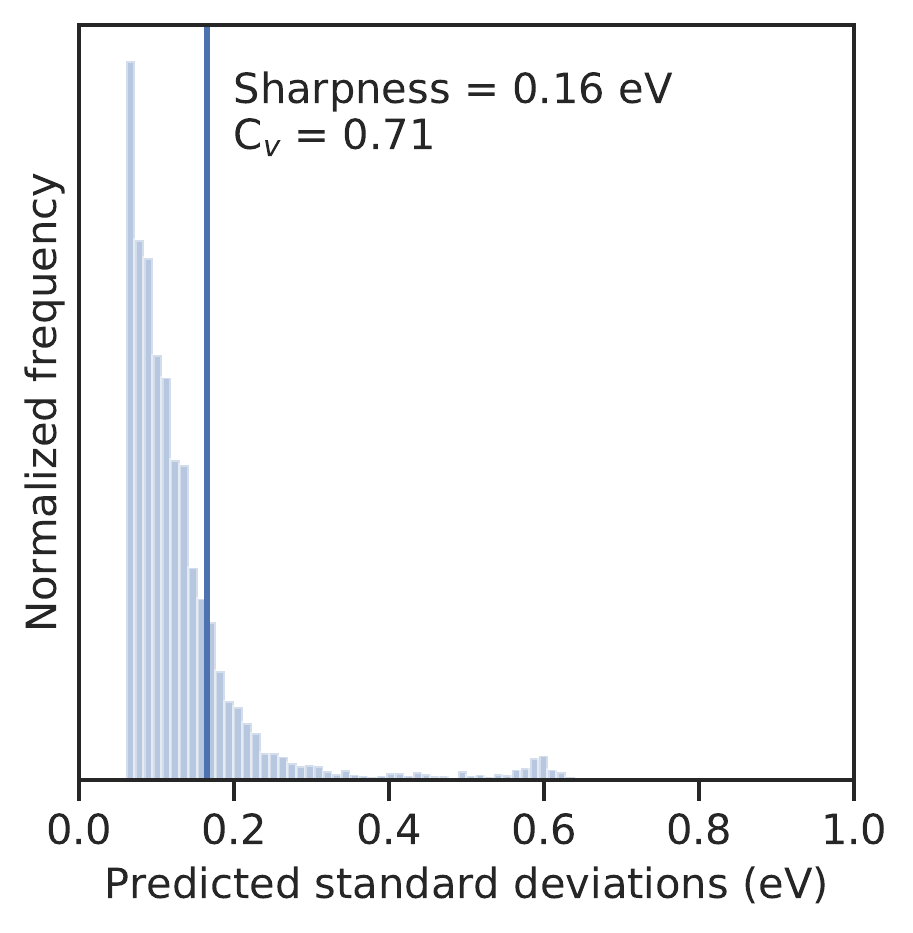}
        \caption{\gls{dNN}}\label{fig:sharpness_dnn}
    \end{subfigure}
    \begin{subfigure}{0.32\textwidth}
        \includegraphics[width=\textwidth]{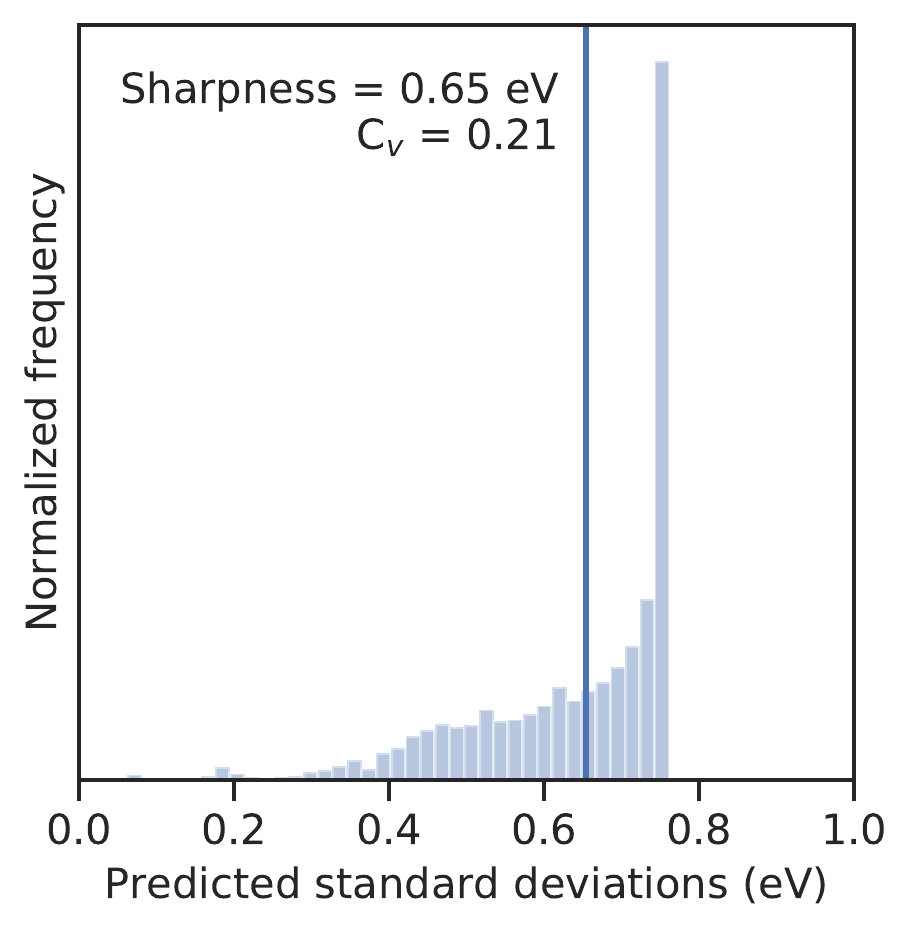}
        \caption{\gls{GP}}\label{fig:sharpness_gp}
    \end{subfigure}
    \begin{subfigure}{0.32\textwidth}
        \includegraphics[width=\textwidth]{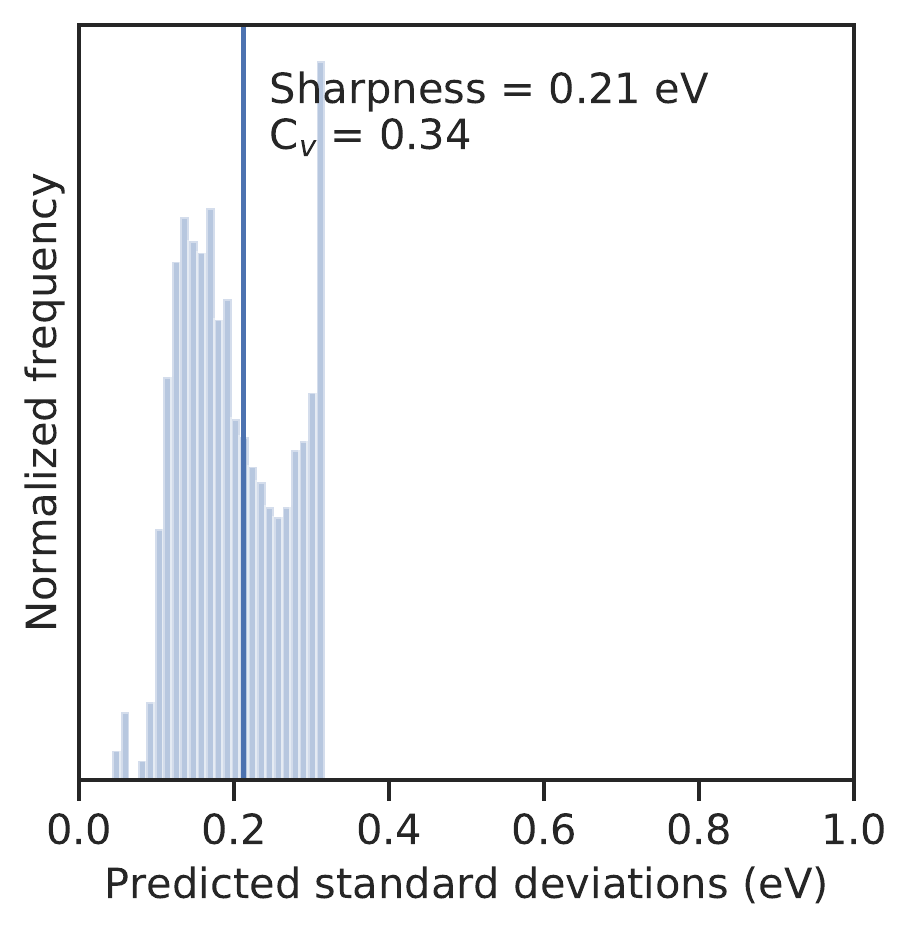}
        \caption{\gls{dGP}}\label{fig:sharpness_dgp}
    \end{subfigure}
    \begin{subfigure}{0.32\textwidth}
        \includegraphics[width=\textwidth]{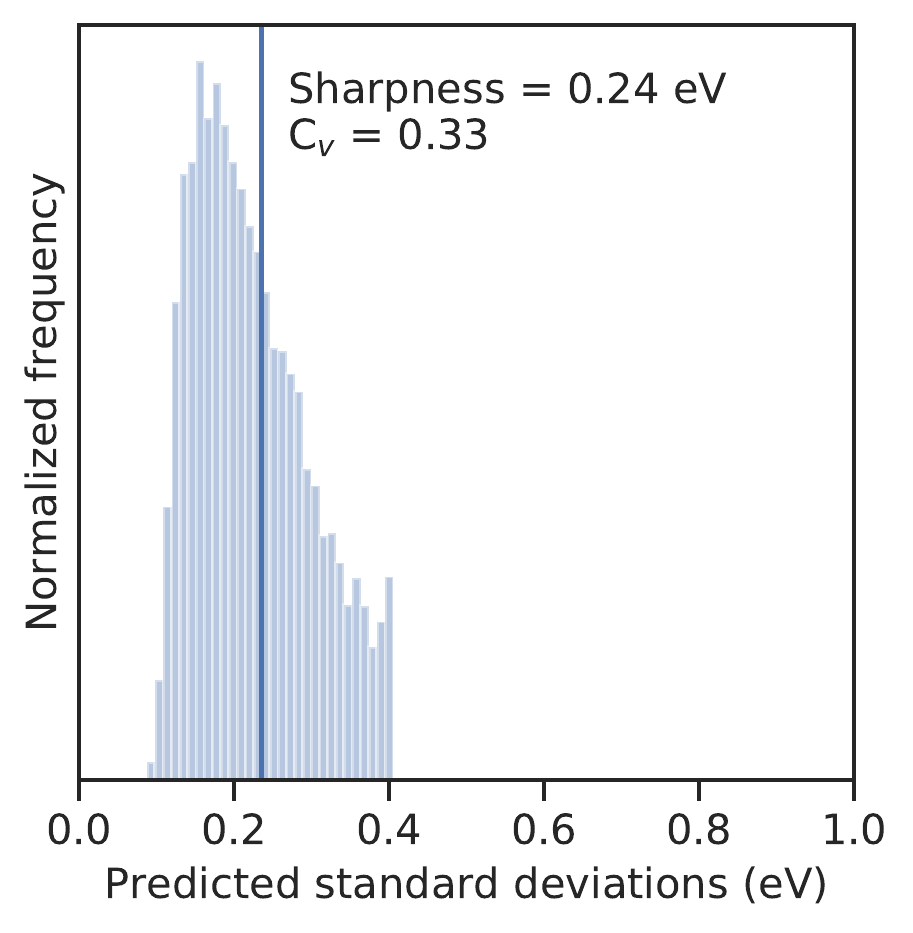}
        \caption{\gls{CFGP}}\label{fig:sharpness_cfgp}
    \end{subfigure}
    \caption{Distribution plots of the \gls{ML}-predicted standard deviations for each method.
    Sharpness values are indicated by vertical lines.}\label{fig:sharpness}
\end{figure}

Figure~\ref{fig:parity} contains parity plots for all \gls{UQ} methods studied here; Figure~\ref{fig:calibration} contains all calibration curves; and Figure~\ref{fig:sharpness} contains all distribution plots of the \gls{ML}-predicted \gls{UQ}s.
These figures illustrate the accuracy, calibration, and sharpness of the different \gls{UQ} methods, respectively.
Table~\ref{tab:results} lists their performance metrics.

\begin{table}
    \centering
    \resizebox{\textwidth}{!}{%
        \begin{tabular}{lrrrrrrrrrr}
            \toprule
            Method              & \gls{MDAE}    & \gls{MAE} & \gls{RMSE}    & \gls{MARPD}   & \gls{R2}  & MisCal    & CalErr    & Sha   & \gls{Cv}    & \gls{NLL}$\cdot$10\textsuperscript{3}\\
            \midrule
            \gls{NN}            & 0.11          & 0.19      & 0.34          & 61            & 0.80      & N/A       & N/A       & N/A   & N/A         & N/A \\
            \gls{NN} ensemble   & 0.11          & 0.18      & 0.32          & 59            & 0.82      & 0.12      & 1.70      & 0.14  & 1.06        & 192.08 \\
            \gls{BNN}           & 0.11          & 0.19      & 0.31          & 59            & 0.83      & 0.20      & 5.32      & 0.03  & 0.30        & 669.61 \\
            \gls{dropout}       & 0.11          & 0.19      & 0.34          & 61            & 0.79      & 0.14      & 2.52      & 0.09  & 0.82        & 7.38$\cdot$10\textsuperscript{14} \\
            \gls{dNN}           & 0.11          & 0.19      & 0.34          & 59            & 0.80      & 0.05      & 0.39      & 0.16  & 0.71        & 18.61 \\
            \gls{GP}            & 0.11          & 0.21      & 0.39          & 61            & 0.73      & 0.14      & 2.35      & 0.65  & 0.21        & 6.41 \\
            \gls{dGP}           & 0.11          & 0.19      & 0.33          & 59            & 0.81      & 0.03      & 0.08      & 0.21  & 0.34        & 6.09 \\
            \gls{CFGP}          & 0.11          & 0.19      & 0.33          & 59            & 0.80      & 0.03      & 0.13      & 0.24  & 0.33        & 2.80 \\
            \bottomrule
        \end{tabular}
    }
    \caption{Performance metrics for all methods used in this study, which include: \glsentryfull{MDAE}, \glsentryfull{MAE}, \glsentryfull{RMSE}, \glsentryfull{MARPD}, \glsentryfull{R2}, miscalibration area (MisCal), calibration error (CalErr), sharpness (Sha), \glsentryfull{Cv}, and \glsentryfull{NLL}.
    The units of \gls{MDAE}, \gls{MAE}, \gls{RMSE}, and sharpness are all in \gls{eV}.
    The units of \gls{MARPD} are in \%.
    The miscalibration area, calibration error, \gls{Cv}, and \gls{NLL} are unitless.}\label{tab:results}
\end{table}

Regarding accuracy:  All methods' \gls{MDAE} results are virtually identical, and their \gls{MAE} results are within 10\% of each other.
This suggests that all methods have comparable predictive accuracies for inliers.
The plain \gls{GP} has a higher \gls{RMSE} value than the rest of the methods, indicating that it has the worst predictive accuracy for outliers.
Correlations between residuals and uncertainty estimates are discussed in the Supplementary Information section briefly.

Regarding calibration:  The \gls{NN} ensemble, \gls{BNN}, and \gls{dropout} are overconfident; the \gls{GP} is underconfident; and the \gls{dNN}, \gls{dGP}, and \gls{CFGP} models are relatively calibrated.
The three more calibrated methods all share a characteristic that the other methods do not:  They all start with a \gls{NN} that is dedicated for prediction alone, and then they end with some other in-series method to estimate uncertainty.
Interestingly, this in-series method of learning predictions and then learning uncertainties is similar in spirit to how gradient boosted models ``learn in stages'' using an ensemble of models.

Regarding sharpness:  The \gls{NN} ensemble, \gls{BNN}, and \gls{dropout} models yield the most sharp uncertainties, although they do so at the cost of calibration.
Among the three more calibrated models, the \gls{dNN} yields the lowest sharpness of 0.16 eV while the \gls{dGP} and \gls{CFGP} yield sharpnesses of 0.21 and 0.24 eV, respectively.
Note how \gls{GP}-based \gls{UQ} methods tend to yield less sharp uncertainties than methods based purely on \gls{NN}s.
This suggests that \gls{GP}s may yield more conservative \gls{UQ}s.

Regarding \gls{NLL}:  The \gls{CFGP} method yields the best (i.e., lowest) \gls{NLL} value of \textit{ca.} 2,800 while both the \gls{GP} and \gls{dGP} models yield relatively moderate \gls{NLL} values of \textit{ca.} 6,000.
Note how the under-confident \gls{GP} model has a worse miscalibration area, calibration error, and sharpness than the \gls{dNN} but a better \gls{NLL} value.
Simultaneously, the three most over-confident and sharp models (\gls{NN} ensemble, \gls{BNN}, and \gls{dropout}) yield the worst \gls{NLL} results.
This shows that better \gls{NLL} values correlate with relatively conservative estimates of \gls{UQ}, but not with relatively liberal estimates.
In other words:  If we use \gls{NLL} as our main performance metric, then we will favor under-confident \gls{UQ} estimates in lieu of over-confident estimates.

Given the performance metrics for accuracy, calibration, sharpness, and \gls{NLL}, we expect the \gls{CFGP} or \gls{dGP} methods to yield the best performing \gls{UQ} models for our dataset.
When choosing \gls{UQ} methods for different applications, other factors should be considered.
For example:  Although the \gls{dGP} method performed relatively well, it relied on hand-crafted features.
If future researchers wish to use the \gls{dGP} method to predict other properties from atomic structures, they may have to define their own set of features.
This process of feature engineering is non-trivial and varies from application to application.
In some cases, it may be easier to use a \gls{UQ} method that does not require any additional features beyond the \gls{NN} input, such as \gls{dNN} or \gls{CFGP}.
This is why declare \gls{CFGP} as the method of choice for our study here; it has a relatively competitive accuracy, calibration, and sharpness while requiring less information than \gls{dGP}.

Another factor to consider is the overhead cost of implementation.
For example:  The \gls{NN} ensemble method is arguably the simplest \gls{NN}-based \gls{UQ} method used here and may be the easiest method to implement.
Conversely, \gls{NN} ensembles also have a higher computational training cost than some of the other methods used here, such as \gls{dNN} or \gls{CFGP}.
This high training cost is exacerbated if the ensemble is meant to be used in an active framework where the model needs to be trained continuously.
As another example:  The \gls{BNN} method yielded perhaps the worst results of all the methods studied here.
It could be argued that further optimization of the \gls{BNN} could have resulted in higher performance.
But creation and training of \gls{BNN}s is still an active area of research with less literature and support than \gls{GP}s or non-Bayesian \gls{NN}s.
This lack of support led to us spending nearly twice as long creating a \gls{BNN} compared to the other methods.
It follows that further optimization of the \gls{BNN} would be non-trivial and may not be worth the overhead investment.


\section{Conclusions}

We examined a procedure for comparing different methods for \glsentryfull{UQ}.
This procedure considers the accuracy of each method, the honesty of their uncertainty estimates (i.e., their calibration), and the size of their uncertainty estimates (i.e., their sharpness).
To assess accuracy, we outlined a common set of error metrics such as \gls{MAE} or \gls{RMSE}, among others.
To assess calibration, we showed how to create, interpret, and quantify calibration curves.
To assess sharpness, we showed how to calculate and plot sharpness.
To assess all three aspects simultaneously, we suggest using the \glsentryfull{NLL} as a performance metric.
The ensemble of all these metrics and figures can be used to judge the relative performance of various \gls{UQ} methods in a holistic fashion.

As a case study, we tested six different methods for predicting \glsentryfull{DFT} calculated adsorption energies with \gls{UQ}.
The best performing method was a \glsentryfull{CFGP}, which used a pre-trained convolutional output from a \gls{NN} as features for a subsequent \gls{GP} that made probabilistic predictions.
Our studies also showed that the \gls{GP}-based methods we tested tended to yield higher and more conservative uncertainty estimates than the methods that used only \gls{NN}s and \gls{NN} derivatives.
We also found that in-series methods tended to yield more calibrated models---i.e., methods that use one model to make value predictions and then a subsequent model to make uncertainty estimates were more calibrated than models that attempted to make value and uncertainty predictions simultaneously.
These results are limited to our dataset.
Results may vary for studies with different applications, different models, or different hyperparameters.
But the underpinning procedure we used to compare these models is still broadly applicable.

Note that it would be possible to recalibrate\cite{Kuleshov2018} each of the models in this study to improve their uncertainty estimates.
We purposefully omitted recalibration in this study to (1) simplify the illustration of the \gls{UQ} assessment procedure; (2) assess the innate performance of each of these \gls{UQ} methods without confounding with recalibration methods; and (3) reduce overhead investment.
Future work should consider recalibration if the feasible \gls{UQ} methods provide insufficiently calibrated uncertainty predictions

Future work may also consider inductively biased \gls{UQ}s.
For example:  If we used the \gls{BEEF},\cite{Wellendorff2012} then our \gls{DFT} calculated adsorption energies would have been distributions rather than than single point estimates.
Such distributions could be propagated to certain \gls{UQ} surrogate models, e.g., as a variable-variance kernel in a \gls{GP}-type method.
As another example of inductively biased \gls{UQ}s:  A model may be able to make low-uncertainty predictions on a \gls{DFT}-optimized structure and then also make high-uncertainty predictions on a similar but \gls{DFT}-unoptimized structure.
\gls{UQ}s do not need to be derived strictly from data.
They may also be derived from previous knowledge.


\section*{Author information} Corresponding author email:  zulissi@andrew.cmu.edu.
The authors declare no competing financial interest.

\section*{Acknowledgments} This research used resources of the National Energy Research Scientific Computing Center, a DOE Office of Science User Facility supported by the Office of Science of the U.S. Department of Energy under Contract No. DE-AC02-05CH11231. 
We also acknowledge the exceptional documentation and clear examples in the GPyTorch\cite{Gardner2018} repository, which formed the basis on much of the \gls{GP} code used for this work.

\section*{Code availability} Visit \texttt{https://github.com/ulissigroup/uncertainty\_benchmarking} for the code used to create the results discussed in this paper.
The code dependencies are listed inside the repository.

\section*{Data availability} The data that support the findings of this study are available from the corresponding author upon reasonable request.


\clearpage
\bibliography{uncertainty_benchmarking}

\end{document}

\begin{figure}
    \centering
    \includegraphics[width=\textwidth]{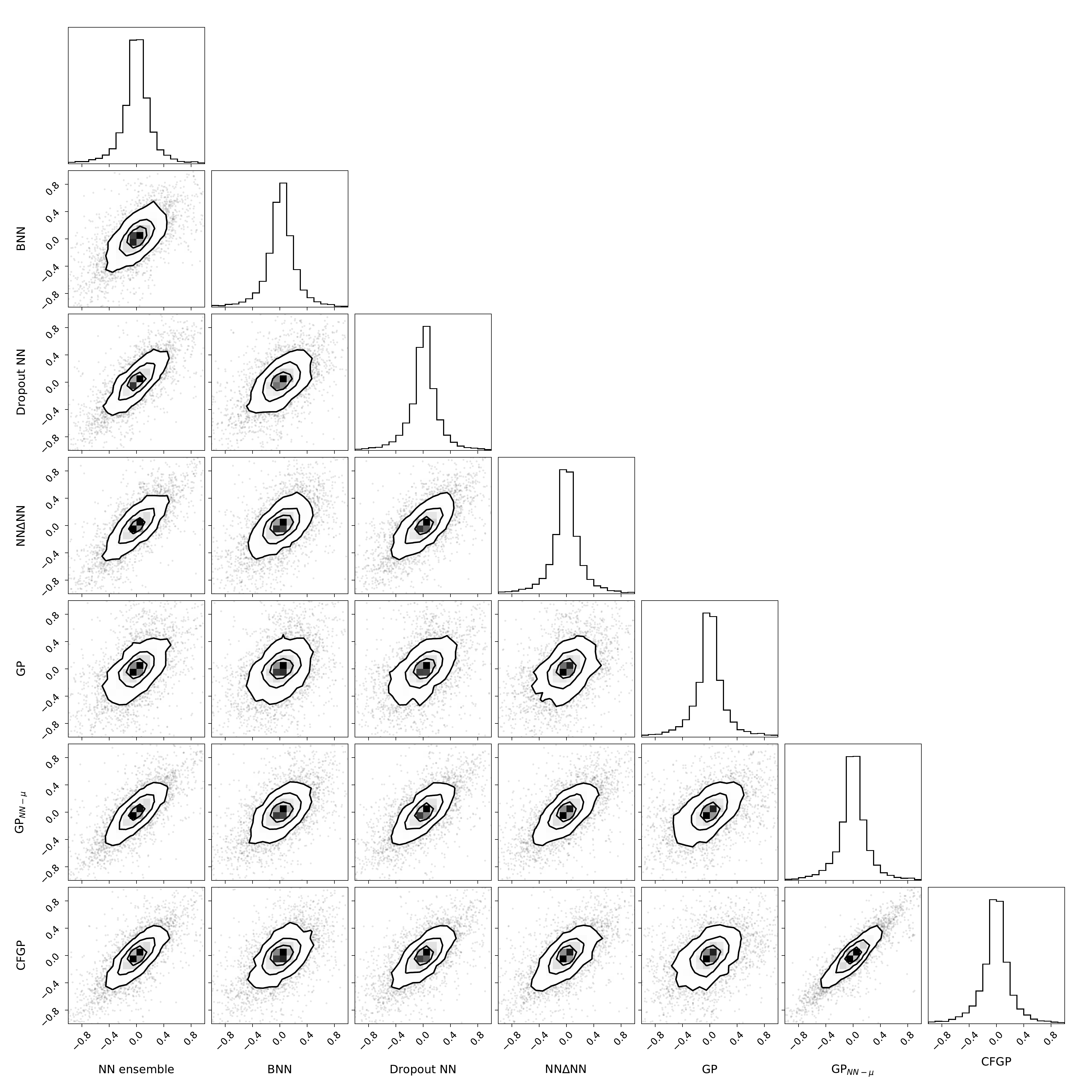}
    \caption{Corner plot of the residuals of all models.
        Each subfigure shows the parity between the residuals of pairs of models.
        Solid contour lines delineate quartiles of the point distribution.
        Single, faded points indicate parity points in the fourth, least dense quartile of points.
        Shaded pixels indicate the highest density of points with darker shading indicating a higher density.
        The figures along the diagonal show histogram distributions of the residuals for each model.
        All units are in eV.
        }\label{fig:resids}
\end{figure}

Figure~\ref{fig:resids} shows how the residuals of each model are correlated with each other.
All pairs of models show a positive correlation between each other.
This suggests that poor predictions made by one model were also made by most other models, qualitatively speaking.
This observation is consistent with the finding that the accuracy of all models in this study were comparable.
No single model was substantially better at predicting the outlying points than any other model.

\begin{figure}
    \centering
    \includegraphics[width=\textwidth]{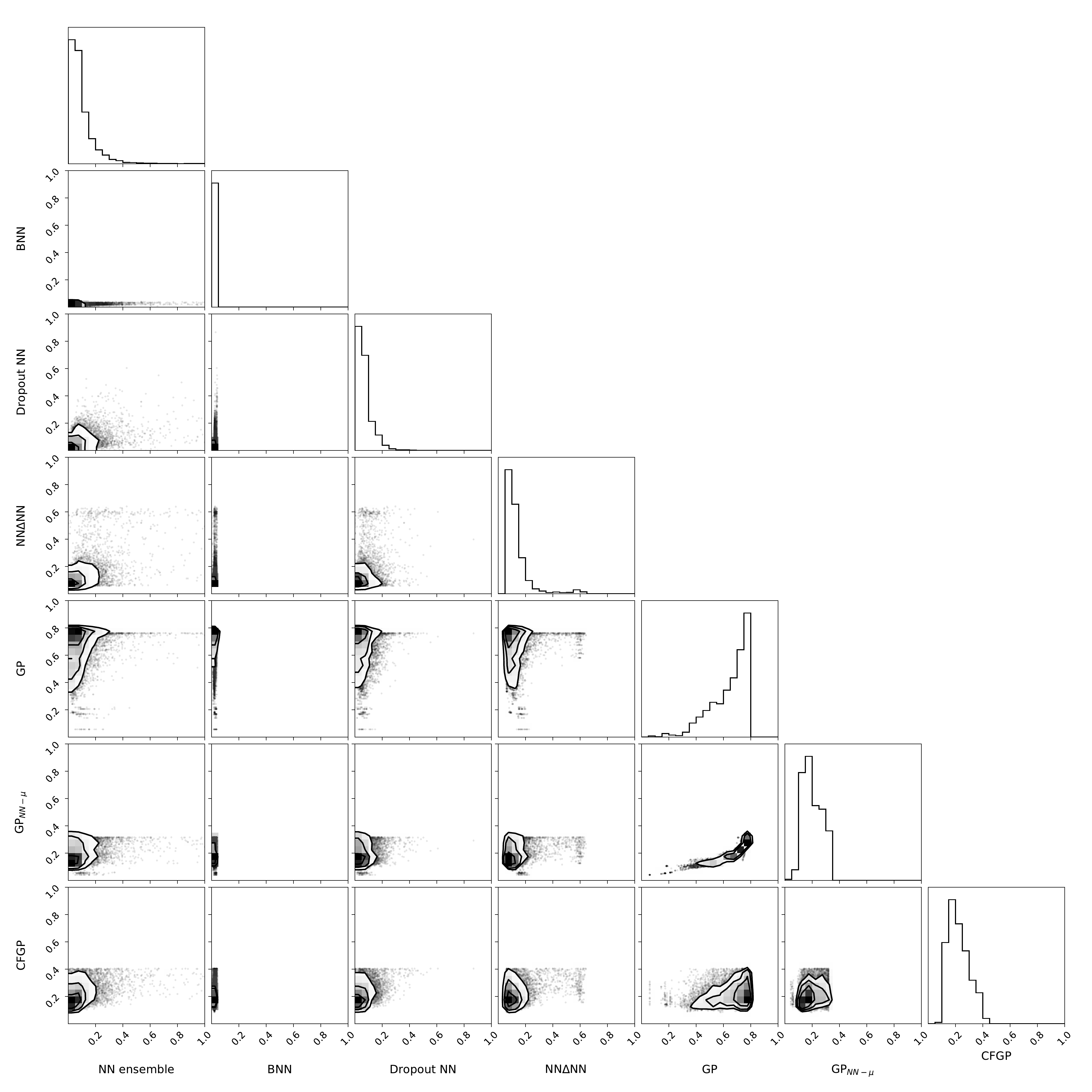}
    \caption{Corner plot of the residuals of all models.
        Each subfigure shows the parity between the estimated standard deviations of pairs of models.
        Solid contour lines delineate quartiles of the point distribution.
        Single, faded points indicate parity points in the fourth, least dense quartile of points.
        Shaded pixels indicate the highest density of points with darker shading indicating a higher density.
        The figures along the diagonal show histogram distributions of the predicted standard deviations for each model.
        All units are in eV.
        }\label{fig:stdevs}
\end{figure}

Figure~\ref{fig:stdevs} shows how the predicted uncertainties of each model are correlated with each other.
The only pattern we could discern was the correlation between the GP and GP$_{NN-\mu}$ methods.
This correlation likely due to the fact that both methods used the same exact feature space for their GPs.
The only difference between the two were their mean functions.

